\documentclass[preprint,12pt,number,nopreprintline]{elsarticle}

\usepackage{amsmath,amssymb,amsfonts}
\usepackage{bm}
\usepackage{graphicx}
\usepackage{subcaption}
\usepackage{array}
\usepackage{booktabs}
\usepackage{siunitx}
\usepackage{multirow}
\usepackage{tabularx}
\usepackage{hyperref}
\hypersetup{hidelinks}
\usepackage{xurl}
\usepackage{cleveref}
\usepackage{xcolor}
\usepackage{placeins}
\usepackage{float}

\newcolumntype{Y}{>{\centering\arraybackslash}X}

\newcommand{\dd}{\mathrm{d}}
\newcommand{\vect}[1]{\bm{#1}}

\newcommand{\OmegaVec}{\bm{\Omega}}

\begin{document}

\begin{frontmatter}

% =========================
% Title information
% =========================
\title{Free-Molecular Face-Flux Preprocessing for Reduced Neutral-Continuity Modeling in
Hall Thrusters: Particle-Based Reference and Deterministic
\texorpdfstring{$S_N$}{SN}-DFEM Realization}

\author[inst1]{Yingjie Chen}
\author[inst1]{Xi Chen}
\author[inst1]{Yinjian Zhao}
\address[inst1]{School of Energy Science and Engineering, Harbin Institute of Technology,
Harbin 150001, People's Republic of China}

\begin{abstract}
Neutral gas transport directly affects the ionization source, propellant utilization, and
low-frequency discharge oscillations in Hall thrusters. High-fidelity particle-based
neutral models or DSMC methods can describe rarefied gas transport, but they are
computationally expensive; in contrast, reduced neutral-continuity models are cheaper but
require a closure for the neutral velocity or face-normal flux. Under a low-pressure
collisionless approximation, this work adopts a free-molecular preprocessing strategy to
provide a reference density field and the mean-velocity or face-normal-flux closure used
by the reduced neutral-continuity equation in a manner consistent with the underlying
transport model.On this basis, a particle-based free-molecular face-flux preprocessor is used as a
stochastic reference, and an $S_N$-DFEM deterministic free-molecular preprocessor is
proposed to generate the corresponding reference density, velocity moments, and
face-normal fluxes within a unified free-molecular transport framework. Results show that
the $S_N$-DFEM preprocessor preserves the main neutral-density and velocity structures and reduces the statistical
error in face-flux closure by about three orders of magnitude in the baseline
continuity-recovery test. A prescribed moderate ionization-loss case further demonstrates
the extension of the framework to free-molecular preprocessing with volumetric neutral
removal.
\end{abstract}


\end{frontmatter}

% =========================
% 1. Introduction
% =========================

\section{Introduction}

Hall thrusters rely on the ionization of neutral propellant atoms and the subsequent
acceleration of ions by an axial electric field \cite{TaccognaLatestprogressi2019}. The
neutral atom distribution governs the ionization source, propellant residence time,
propellant utilization, and low-frequency breathing oscillations. Neutral depletion and
refilling dynamics are central to breathing-mode behavior, while the position and strength
of the ionization source affect discharge current, wall loading, and overall thruster
operation \cite{PetronioStudyofthebreat2024,CaoInfluenceofneut2023}. The near-wall
neutral distribution can also influence local ionization and charge-exchange processes,
thereby modifying plasma-wall interaction and erosion-relevant particle fluxes
\cite{DingEffectsofthemag2017}. An accurate and efficient representation of neutral
transport is therefore important for predictive Hall-thruster simulations.

A wide range of neutral models has been used in Hall-thruster simulations. Some kinetic
plasma simulations prescribe the neutral density as a fixed uniform background
\cite{Croes2Dparticleincel2017} or as a one-dimensional axial profile
\cite{MarnCebrinNonMaxwellianel2024}, thereby eliminating neutral dynamics from the
calculation. Particle-based models retain more kinetic information: collisionless
particle tracking is suitable in the free-molecular limit, while Monte Carlo collision
and DSMC treatments can be added for plasma-neutral reactions and neutral-neutral
rarefied-gas collisions \cite{MiyasakaParticlesimulat2008,GueroultPitfallsinModel2018}.
Hybrid PIC-fluid approaches and precomputed neutral fields provide intermediate
strategies for reducing cost while retaining part of the neutral response
\cite{PanelliAxisymmetricHyb2021,SommierSimulatingPlasm2007,DingEffectsofthemag2017}.
Pan et al.\ \cite{PanPracticalanalys2023} compared fixed-neutral, DSMC,
collisionless-neutral, and fluid-type neutral treatments for a 1-kW SPT-100 Hall
thruster, illustrating how the neutral algorithm itself can affect the predicted
discharge behavior. Recent Hall-thruster simulations have also emphasized the need to
balance model fidelity, transport physics, and computational cost when describing coupled
plasma, neutral, and heavy-species dynamics
\cite{ZhaoReview3DPIC,ShashkovPST2023,YangPST2018,ChenPST2023}.

Within this broad spectrum, reduced neutral-continuity models are attractive because of
their low cost, but their predictive value depends critically on how the neutral velocity
or face-normal flux is closed. The continuity equation provides a conservation law for
neutral density, but it must be supplemented by either the volumetric flux
($\boldsymbol{\Gamma}=n\boldsymbol{u}$) or finite-volume face fluxes ($\Gamma_f$). In
many reduced models, this closure is supplied by prescribed axial velocities, empirical
velocity profiles, or other local transport assumptions
\cite{PetronioStudyofthebreat2024,FarajiEffectsoftheneu2023}. Such treatments are
computationally efficient, but they separate the velocity or flux closure from the
underlying nonlocal transport, especially when inlet injection, geometric shadowing, wall
reflection, and open-boundary escape all influence the local flux.

In low-pressure Hall-thruster channels and near-plume regions, the neutral mean free path
can be comparable to or larger than the characteristic geometric scale. Neutral-neutral
collisions are then infrequent, and transport is governed primarily by inlet injection,
ballistic motion, wall reflection, ionization loss, and boundary escape. Free-molecular
assumptions have been applied to neutral transport in channel-geometry studies
\cite{MaNumericalsimula2020}, and the collisionless neutral algorithm examined by Pan et
al.\ \cite{PanPracticalanalys2023} is explicitly based on this regime. Under such
conditions, the neutral velocity or face-normal flux should not be treated simply as a
local function of density. Instead, it should be obtained from the same geometric and
boundary model that defines the free-molecular transport problem. This motivates a
free-molecular preprocessing strategy: solve a neutral free-molecular transport problem
first, and then use the resulting reference density, mean velocity, and face-normal fluxes
to close a reduced neutral-continuity equation.

A particle-based free-molecular preprocessor provides a direct stochastic construction of
such a closure. Particle trajectory sampling naturally represents inlet injection,
ballistic motion, wall reflection, and open-boundary escape. Residence-time statistics
provide cell-centered density estimates, while face-crossing statistics provide
finite-volume face-normal fluxes. These statistics can be used to define a conservative
face-flux handoff for a reduced neutral-continuity equation. However, velocity moments and
face-normal fluxes are typically more sensitive to particle sampling noise than density
alone. Particle-based neutral models have also been recognized as a source of unphysical
statistical fluctuations that may contaminate the analysis of plasma physics
\cite{KatzNeutralgasfree2011}. In the present work, the particle-based free-molecular
face-flux preprocessor is therefore used as a stochastic reference for the handoff and as
a definition of the target mean-velocity and face-normal-flux closure quantities; the next
step is to generate the corresponding density, velocity moments, and face-normal fluxes
deterministically.

Free-molecular preprocessing and deterministic neutral transport have important
precedents. Katz and Mikellides developed a view-factor-based deterministic
free-molecular neutral algorithm that advances source-surface density contributions using
precomputed geometry-dependent velocities or flux coefficients, producing neutral density
fields without particle statistical noise \cite{KatzNeutralgasfree2011}. Related
approaches have been used in Hall-thruster simulation frameworks such as Hall2De
\cite{Ortega2DFluidPIC2023}, and Araki extended the view-factor model to streaming and
non-Maxwellian inflow conditions for Hall-effect-thruster plume neutrals
\cite{ArakiExtensionViewFactor2019}. Direct kinetic methods provide another deterministic
route by solving the distribution function on a phase-space grid rather than sampling
particles \cite{RaisanenTwodimensionalh2019,HaraOnedimensionalh2012}. These studies show
that free-molecular preprocessing and deterministic neutral transport are established
directions. The present work therefore does not claim the first free-molecular
preprocessor or the first deterministic neutral model; instead, it develops a
complementary $S_N$-DFEM realization and assesses its use for reduced-continuity
face-flux closure.

The deterministic preprocessor developed here solves the collisionless neutral Boltzmann
transport equation in a two-dimensional axisymmetric $r$-$z$ geometry. Molecular speed is
discretized into speed groups, while velocity direction is represented by a
discrete-ordinates, or $S_N$, angular quadrature. For each discrete velocity, the
free-molecular transport equation is solved on the axisymmetric mesh using an upwind
discontinuous finite-element method. Inlet injection, open outflow, and mixed
diffuse/specular wall reflection are treated within the same free-molecular transport
model. Unlike view-factor/source-surface formulations that reduce free-molecular transport
to source contributions and geometric coefficients, the present formulation retains the
discrete angular distribution as a resolved unknown on the mesh. Density, mean velocity,
and face-normal flux moments can therefore be reconstructed from the same angular-flux
solution. Retaining angular information also enables the effects of mixed wall reflection,
angular quadrature, speed-group resolution, and the $S_N$ ray effect to be examined
\cite{AdamsFastiterativeme2002,MercimekDiscontinuousfi2014,CamminadyRayeffectmitiga2019}.

These elements define the free-molecular face-flux preprocessing framework assessed in
this work.

The paper is organized as follows. Section~2 describes the physical model, the
particle-based free-molecular face-flux preprocessor, the deterministic $S_N$-DFEM
preprocessor, boundary conditions, and the flux-based continuity handoff. Section~3
presents the baseline comparison, continuity recovery, resolution studies, wall-reflection
effects, cylindrical-geometry effects, parametric scans, and the prescribed ionization-loss
test. Section~4 summarizes the main conclusions, limitations, and possible extensions.

\section{Method}
\label{sec:method}

This section first summarizes the neutral transport model and its role in the overall
Hall-thruster neutral treatment. The main objective is to avoid relying solely on an
empirical neutral-velocity prescription by using a precomputed free-molecular transport
field that contains the effects of geometry, inlet injection, open boundaries, and wall
reflection. The particle-based FM and deterministic preprocessors are then described as
two numerical ways of generating this precomputed field.

\subsection{Physical Model}
\label{subsec:physical_model}

The physical configuration is a two-dimensional axisymmetric representation of the
Hall-thruster channel and near-plume region in the $(r,z)$ plane. Neutral propellant is
supplied from the anode-side inlet, moves through the channel, is partly redirected by
solid walls, and can escape through open boundaries. This work focuses on the
large-Knudsen-number limit, where neutral-neutral collisions are neglected over the
geometric scale of interest and neutral atoms are not directly accelerated by
electromagnetic fields.

The present model is two-dimensional in physical space and uses velocity directions in
the meridional $r$-$z$ plane. It is therefore a reduced axisymmetric free-molecular
model rather than a full 2D3V neutral kinetic description.

In this limit, the neutral motion is determined primarily by geometry and boundary
interactions. After emission from the inlet or a wall, a neutral atom travels
ballistically until it reaches another boundary. The inlet specifies the incoming
neutral flux and velocity distribution. Open boundaries remove outgoing neutrals and do
not inject neutrals from outside the domain. Solid walls are modeled by a mixed
reflection law: a fraction of the incident flux is reflected specularly, while the
remaining fraction is diffusely re-emitted according to the wall temperature.

The axisymmetric geometry is retained through cylindrical cell volumes and face areas,
including the cell volume $V_i=\int_{K_i}2\pi r\,dr\,dz$, radial-face areas $A_r=2\pi
r_f\Delta z$, and axial-face areas $A_z=\pi(r_{i+1/2}^2-r_{i-1/2}^2)$. These geometric
factors are responsible for the annular conservation behavior discussed in
\cref{subsubsec:cylindrical_geometry_effect}. The preprocessor is therefore used to
compute the neutral transport structure implied by this physical model before the plasma
calculation is advanced. Its main outputs are the reference neutral density, mean
velocity, and face-normal number fluxes. The face-normal fluxes are used directly in the
finite-volume continuity handoff described below.

The primary task of the preprocessor is to generate the free-molecular transport field
implied by the inlet, open-boundary, wall-reflection, and cylindrical-geometry models.
As an application and consistency check, the reconstructed face-normal flux moments are
also passed to a finite-volume neutral-continuity update. In the finite-volume
implementation used below, this handoff is performed through face-normal number fluxes
rather than only through cell-centered velocity moments; the corresponding closure is
described in
\cref{subsec:face_flux_continuity_closure}.

\subsection{Particle-Based Free-Molecular (FM) Face-Flux Preprocessor}
\label{subsec:particle_preprocessor}

Free-molecular neutral preprocessing has been used in reduced Hall-thruster neutral
models to separate the fast geometric transport calculation from the subsequent
neutral-density update. In this approach, a collisionless neutral calculation is first
performed on the fixed thruster geometry. The resulting steady transport pattern is then
passed to a reduced continuity equation instead of tracking neutral particles during
every plasma time step.

The particle-based free-molecular (FM) preprocessor used here follows this idea, but
records both cell residence statistics and face-crossing statistics during a
statistically steady sampling window. Between boundary events, each particle satisfies
the collisionless free-flight equations
\begin{equation}
    \frac{\dd \mathbf{x}_p}{\dd t}
    =
    \mathbf{v}_p,
    \qquad
    \frac{\dd \mathbf{v}_p}{\dd t}
    =
    \mathbf{0}.
    \label{eq:particle_free_flight}
\end{equation}
When a particle reaches a solid wall, the outgoing velocity is selected from the same
mixed wall model used throughout this work. With probability $1-\alpha_d$, the velocity
is specularly reflected,
\begin{equation}
    \mathbf{v}_p^{+}
    =
    \mathbf{v}_p^{-}
    -
    2
    \left(
    \mathbf{v}_p^{-}\cdot\hat{\mathbf n}_w
    \right)
    \hat{\mathbf n}_w ,
    \label{eq:particle_specular_reflection}
\end{equation}
where $\hat{\mathbf n}_w$ is the local wall normal and the superscripts $-$ and $+$
denote incoming and outgoing velocities. With probability $\alpha_d$, the particle is
diffusely re-emitted from a wall-temperature half-range Maxwellian over the gas-side
outgoing half-space. Open boundaries remove outgoing particles from the domain.

If $T_{\mathrm{avg}}$ is the averaging time and $w_p$ is the macro-particle weight, the
reference density in cell $K_i$ is estimated from the residence time as
\begin{equation}
    n_i^{\mathrm{ref}}
    =
    \frac{1}{V_i T_{\mathrm{avg}}}
    \sum_p
    w_p\,\Delta t_{p,i},
    \label{eq:particle_residence_density}
\end{equation}
where $\Delta t_{p,i}$ is the time spent by particle $p$ in cell $K_i$ during the
averaging interval. For an oriented internal face $f$, the signed face-normal
number-flux density is estimated from crossing events,
\begin{equation}
    \Gamma_f^{\mathrm{ref}}
    =
    \frac{1}{A_f T_{\mathrm{avg}}}
    \sum_{p\in\mathcal C_f}
    s_{p,f} w_p ,
    \label{eq:particle_face_crossing_flux}
\end{equation}
where $\mathcal C_f$ is the set of crossings of face $f$ and $s_{p,f}=+1$ or $-1$
denotes whether the crossing follows or opposes the chosen face orientation.
Open-boundary outgoing fluxes are recorded in the same way.

For the reduced-continuity handoff used in this study, the crossing estimate in
\cref{eq:particle_face_crossing_flux} is used as the reference face transport
quantity.

This construction makes the handoff consistent with the finite-volume continuity update.
The continuity solver uses the measured face flux as the reference transport quantity
and constructs an upwind face velocity from the reference density only when a velocity
is needed by the numerical flux. Therefore, when the density equals the preprocessed
reference density, the finite-volume internal flux recovers the measured free-molecular
face flux. This face-level consistency is used below as a reduced-continuity
verification rather than as a change to the underlying free-molecular wall or inlet
physics.

\subsection{\texorpdfstring{Deterministic $S_N$--DFEM Preprocessor}{Deterministic SN--DFEM Preprocessor}}
\label{subsec:sn_solver}

The deterministic preprocessor uses a discrete-ordinates ($S_N$) discontinuous
finite-element method (DFEM) formulation. It solves the same collisionless
free-molecular neutral transport problem as the particle-based FM preprocessor. Instead
of particle sampling, it uses deterministic quadrature in velocity space and an upwind
discontinuous finite-element discretization in physical space. Thus, both preprocessors
use the same inlet source, open-boundary treatment, mixed wall reflection model, and
output definitions; they differ only in how the free-molecular transport field is
generated.

The molecular velocity is represented by a speed group and an angular ordinate,
\begin{equation}
    \mathbf{v}_{g,a}
    =
    v_g\OmegaVec_a,
    \qquad
    \OmegaVec_a=(\mu_a,\eta_a),
\end{equation}
where $g=1,\ldots,N_g$ and $a=1,\ldots,N_\Omega$. The discretization follows the
discrete-ordinates transport idea, but the present neutral implementation uses speed
groups directly rather than conventional energy groups
\cite{LewisMillerComputational1984,AdamsFastiterativeme2002}. For a fixed neutral
species, the kinetic energy $E=m_nv_g^2/2$ is in one-to-one correspondence with the speed,
so the speed-group notation is equivalent to an energy grouping but is more convenient
for free-molecular residence time, face fluxes, and absorption attenuation. For each
speed-angle state, the ordinate $\OmegaVec_a$ lies
in the two-dimensional meridional $(r,z)$ plane. The two components $\mu_a$ and $\eta_a$
are therefore the radial and axial direction cosines of the same angular ordinate,
satisfying $\mu_a^2+\eta_a^2=1$; they do not represent two independent angular
variables. The baseline collisionless transport equation is
\begin{equation}
    \mu_a\frac{\partial\psi_{g,a}}{\partial r}
    +
    \eta_a\frac{\partial\psi_{g,a}}{\partial z}
    =0 .
    \label{eq:sn_discrete_transport}
\end{equation}
Here $\psi_{g,a}$ is the phase-space density associated with speed $v_g$ and ordinate
$\OmegaVec_a$. The speed-group quadrature and inlet-flux normalization are given in
\cref{appsubsec:speed_inlet}.

For each $(g,a)$, Eq.~\eqref{eq:sn_discrete_transport} is solved on the cylindrical
$r$--$z$ mesh by a standard upwind DFEM transport sweep
\cite{ReedHill1973,MercimekDiscontinuousfi2014}. In compact form, the local weak
statement on a cell $K$ is
\begin{equation}
    -
    \int_K
    \psi_{g,a}^h
    \OmegaVec_a\cdot\nabla_{r,z}\varphi
    \,\dd V
    +
    \int_{\partial K}
    \varphi
    \psi_{g,a}^{\ast}
    \left(
    \OmegaVec_a\cdot\vect{n}
    \right)
    \,\dd A
    =0,
    \label{eq:sn_weak_form_main}
\end{equation}
where $\dd V$ and $\dd A$ are the cylindrical wedge volume and face-area measures. The
numerical trace $\psi_{g,a}^{\ast}$ is selected by upwinding: it is the current-cell
trace on outgoing faces and the upwind neighboring or boundary trace on incoming faces.
The sweep order follows the signs of $\mu_a$ and $\eta_a$.

The boundary conditions mirror those in the particle preprocessor. The inlet imposes an
incoming distribution normalized to the prescribed number flux, open boundaries impose
zero incoming flux from outside the domain, and solid walls use a mixed diffuse/specular
gas-surface reflection operator. The wall operator maps the outgoing wall trace onto the
incoming wall trace and is written as
\begin{equation}
    \mathcal{R}_w
    =
    (1-\alpha_d)\mathcal{R}_{\mathrm{sp}}
    +
    \alpha_d\mathcal{R}_{\mathrm{df}},
    \label{eq:wall_operator_split}
\end{equation}
where $\alpha_d$ is the diffuse reflection fraction. The specular part maps an outgoing
ordinate to its mirror-reflected incoming preimage, while the diffuse part redistributes
the total outgoing wall flux over the incoming half-space according to the wall
temperature. Because wall-reflected incoming traces depend on the outgoing angular
solution, the directional sweeps are embedded in a boundary-coupled fixed-point
iteration. The wall operator, diffuse normalization, and iteration residual are provided
in \cref{appsubsec:wall_bc,appsubsec:wall_iteration}.

After convergence, cell moments are reconstructed by quadrature over all discrete
velocity states,
\begin{equation}
    n_{i,k}
    =
    \sum_{g=1}^{N_g}
    \sum_{a=1}^{N_\Omega}
    w_{g,a}\overline{\psi}_{g,a,i,k},
    \label{eq:sn_density}
\end{equation}
\begin{equation}
    \Gamma_{r,i,k}
    =
    \sum_{g=1}^{N_g}
    \sum_{a=1}^{N_\Omega}
    w_{g,a}v_g\mu_a\overline{\psi}_{g,a,i,k},
    \qquad
    \Gamma_{z,i,k}
    =
    \sum_{g=1}^{N_g}
    \sum_{a=1}^{N_\Omega}
    w_{g,a}v_g\eta_a\overline{\psi}_{g,a,i,k}.
    \label{eq:sn_flux_components}
\end{equation}
The mean velocities follow from $u_r=\Gamma_r/n$ and $u_z=\Gamma_z/n$ where $n>0$. The
converged angular solution is also used to reconstruct signed internal and open-boundary
face-normal fluxes using upwind traces. These outputs are written in the same layout as
the particle-based FM face-flux preprocessor; the face-normal flux reconstruction used
for the continuity handoff is summarized in
\cref{appsubsec:face_flux_reconstruction}.

\subsection{Face-Flux Continuity Closure}
\label{subsec:face_flux_continuity_closure}

The free-molecular preprocessing step provides a steady reference transport field. In
the reduced neutral model, this reference field is used to close a finite-volume
continuity equation for the neutral density. The key point of the present implementation
is that the handoff is performed through face-normal number fluxes.

Let $n_i^{\mathrm{ref}}$ denote the reference neutral density in finite-volume cell
$K_i$, and let $\Gamma_f^{\mathrm{ref}}$ denote the signed reference number-flux density
on face $f$. The superscript ``ref'' denotes a preprocessed free-molecular field
obtained either from the particle-based FM face-flux preprocessor or from the
deterministic $S_N$ solver. The reduced neutral continuity equation is written as
\begin{equation}
    \frac{\partial n}{\partial t}
    +
    \nabla\cdot\boldsymbol{\Gamma}
    =
    Q_n .
    \label{eq:reduced_neutral_continuity}
\end{equation}
Here $n$ is the evolving neutral number density, $\boldsymbol{\Gamma}$ is the neutral
number-flux field represented through the face-flux closure below, and $Q_n$ denotes a
generic net neutral production or loss term. Detailed ionization or recombination models
are introduced separately. For a finite-volume cell $K_i$, the continuity equation is
advanced by an explicit finite-volume update in pseudo-time:
\begin{equation}
    n_i^{p+1}
    =
    n_i^{p}
    -
    \frac{\Delta t}{V_i}
    \sum_{f\in\partial K_i}
    A_f F_{i,f}^{p}
    +
    \Delta t Q_{n,i}^{p},
    \label{eq:fv_continuity}
\end{equation}
where $p$ indexes the pseudo-time step, $V_i$ is the cell volume, $A_f$ is the face
area, and $F_{i,f}^{p}$ is the outward numerical number-flux density through face $f$
with respect to cell $K_i$.

For an internal face $f$ shared by a left cell $L$ and a right cell $R$, a fixed face
orientation is chosen from $L$ to $R$. The preprocessed reference flux is converted into
an effective face velocity by
\begin{equation}
    u_f
    =
    \frac{
    \Gamma_f^{\mathrm{ref}}
    }{
    n_{\mathrm{ref},f}^{\mathrm{up}}
    },
    \label{eq:effective_face_velocity}
\end{equation}
where the reference upwind density is
\begin{equation}
    n_{\mathrm{ref},f}^{\mathrm{up}}
    =
    \begin{cases}
    n_L^{\mathrm{ref}},
    &
    \Gamma_f^{\mathrm{ref}}>0,
    \\[4pt]
    n_R^{\mathrm{ref}},
    &
    \Gamma_f^{\mathrm{ref}}<0 .
    \end{cases}
    \label{eq:reference_upwind_density}
\end{equation}
For a general evolving density field, the internal-face flux in the chosen face
orientation is then evaluated by the upwind relation
\begin{equation}
    F_f(n)
    =
    u_f n_f^{\mathrm{up}},
    \label{eq:continuity_upwind_flux}
\end{equation}
with
\begin{equation}
    n_f^{\mathrm{up}}
    =
    \begin{cases}
    n_L,
    &
    u_f>0,
    \\[4pt]
    n_R,
    &
    u_f<0 .
    \end{cases}
    \label{eq:continuity_upwind_density}
\end{equation}
If $\Gamma_f^{\mathrm{ref}}=0$, the corresponding face velocity is set to zero. The
oriented face flux $F_f$ is inserted with opposite signs in the two adjacent cell
balances in Eq.~\eqref{eq:fv_continuity}.

This construction recovers the preprocessed transport field at the reference state:
\begin{equation}
    F_f(n^{\mathrm{ref}})
    =
    u_f n_{\mathrm{ref},f}^{\mathrm{up}}
    =
    \Gamma_f^{\mathrm{ref}} .
    \label{eq:reference_flux_recovery}
\end{equation}
Thus, the handoff from the free-molecular preprocessor to the continuity solver is
consistent at the face-flux level.

Open-boundary outflow faces are treated similarly. If $\Gamma_b^{\mathrm{ref}}$ is the
outward reference flux on a boundary face adjacent to cell $c$, then
\begin{equation}
    u_b
    =
    \frac{\Gamma_b^{\mathrm{ref}}}{n_c^{\mathrm{ref}}},
    \qquad
    F_b(n)=u_b n_c .
    \label{eq:boundary_outflow_closure}
\end{equation}
No incoming flux is imposed at open boundaries. The physical inlet is handled
separately: on the prescribed inlet band, the incoming flux is fixed by the inlet model
and is not rescaled by the unknown adjacent-cell density.

Equation~\eqref{eq:fv_continuity} is iterated until the steady density field converges.
This construction provides a consistency check: when the density equals the preprocessed
reference density, the finite-volume internal flux recovers the preprocessed
free-molecular face flux.

% =========================
% 3. Results and Discussion
% =========================
\section{Results and Discussion}
\label{sec:results}

\subsection{Comparison and Verification}
\label{subsec:comparison_verification}

\subsubsection{Baseline Case}
\label{subsubsec:baseline_case}

The baseline case uses the common geometry and boundary model described in
\cref{subsec:physical_model}. Its parameters include the inlet source strength,
inlet temperature, axial drift velocity, wall temperature, diffuse reflection fraction,
and the numerical settings used by the FM and $S_N$ preprocessors. Unless otherwise
stated, the baseline uses the parameter set listed in
\cref{tab:baseline_case_parameters}. The $S_N$ baseline uses
$N_\Omega=400$ angular ordinates and $N_g=16$ speed groups, which is the reference
resolution selected for the physical parameter scans.

\begin{table}[!tbp]
    \centering
    \caption{Baseline physical parameters and numerical settings.}
    \label{tab:baseline_case_parameters}
    \small
    \begin{tabularx}{0.84\linewidth}{@{}lX@{}}
        \toprule
        Quantity & Baseline value \\
        \midrule
        \multicolumn{2}{@{}l}{\textit{Geometry and physical model}} \\
        Channel radial range & $R_{\mathrm{ch}}\in[0.01357,\ 0.02489]\ \mathrm{m}$ \\
        Channel axial length & $Z_{\mathrm{ch}}=7.94\times10^{-3}\ \mathrm{m}$ \\
        Inlet radial width & $\Delta r_{\mathrm{in}}=8.04\times10^{-3}\ \mathrm{m}$ \\
        Inlet reference density & $n_{\mathrm{in}}=5.0\times10^{18}\ \mathrm{m^{-3}}$ \\
        Inlet temperature & $T_{\mathrm{in}}=550\ \mathrm{K}$ \\
        Wall temperature & $T_w=550\ \mathrm{K}$ \\
        Inlet axial drift velocity & $u_{z0}=300\ \mathrm{m\,s^{-1}}$ \\
        Diffuse reflection fraction & $\alpha_d=0.7$ \\
        \midrule
        \multicolumn{2}{@{}l}{\textit{FM sampling setup}} \\
        Injection rate & $500$ numerical particles per step \\
        Macro-particle weight & $w_p=1.65\times10^7$ neutrals per particle \\
        Particle time step and run length & $\Delta t=1.0\times10^{-7}\ \mathrm{s}$, $4800$ steps \\
        Steady-state criterion & $<1\%$ particle-count change over $320$ steps \\
        \midrule
        \multicolumn{2}{@{}l}{\textit{$S_N$ velocity-space resolution}} \\
        Angular ordinates & $N_\Omega=400$ \\
        Speed groups & $N_g=16$ \\
        \bottomrule
    \end{tabularx}
\end{table}

\begin{figure}[!tbp]
    \centering
    \includegraphics[width=0.78\linewidth]{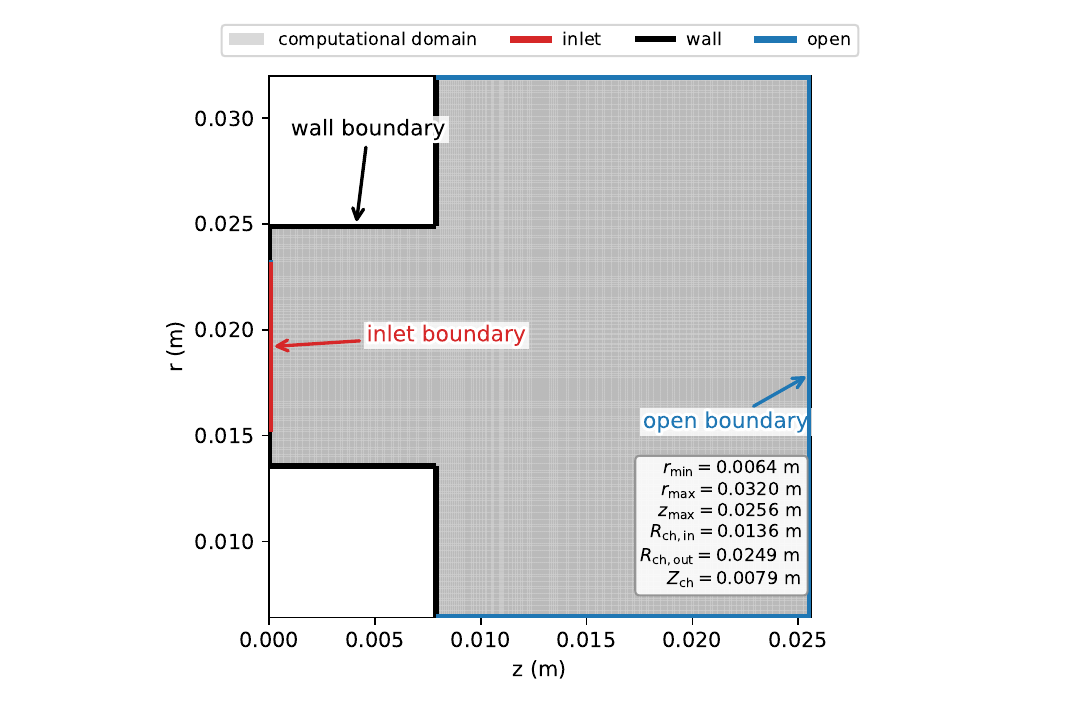}
    \caption{Baseline computational domain and boundary types.}
    \label{fig:results_domain_boundary}
\end{figure}

The two compared preprocessors are the particle-based FM face-flux preprocessor and the
deterministic $S_N$ face-flux preprocessor. This subsection defines the common test case
and comparison objects; field comparisons and error measures are reported in
\cref{subsubsec:comparison}. The baseline domain and boundary classification are shown in
\cref{fig:results_domain_boundary}. The inlet band is located on the anode-side
boundary, open boundaries allow outgoing neutrals to leave the domain, and the remaining
solid surfaces use the mixed diffuse/specular wall-reflection model described in
\cref{subsec:physical_model}.

The figure defines the inlet, open-boundary, and wall-reflection regions used by both
preprocessors and confirms that the two methods share the same geometric and boundary
setup before detailed field comparisons are made.

Because this baseline is formulated in an axisymmetric cylindrical geometry, some radial
asymmetry in the plotted density field is expected from the annular area weighting. This
geometric effect is isolated in the next subsection before the FM and \(S_N\) fields are
compared.

\subsubsection{Cylindrical Geometry Effect}
\label{subsubsec:cylindrical_geometry_effect}

To distinguish cylindrical-geometry effects from numerical asymmetry, a shifted-radius
test is performed. In an axisymmetric \(r\)-\(z\) formulation, radial transport is
affected by the annular area factor \(2\pi r\). Therefore, even under otherwise simple
free-molecular transport, the lower- and higher-radius sides are not geometrically
equivalent.

The shifted-radius case keeps the same logical geometry and radial thickness as the
original plotted active computational domain, but translates all radial coordinates to a
much larger radius. This test is not intended to represent a physical Hall thruster
geometry. Instead, it is used as a numerical radius-scaling diagnostic. For the full
plotted active radial range, the original-radius case has
\(\Delta r/r_{\min}\approx4.0\), so the cylindrical area factor changes appreciably
across the plotted domain. In the shifted-radius case,
\(\Delta r/r_{\min}\approx2.6\times10^{-4}\), so \(2\pi r\) is nearly constant over the
same logical active domain. The geometry-induced radial bias should therefore be
strongly suppressed.

\begin{figure}[!hbp]
    \centering
    \includegraphics[width=0.98\linewidth]{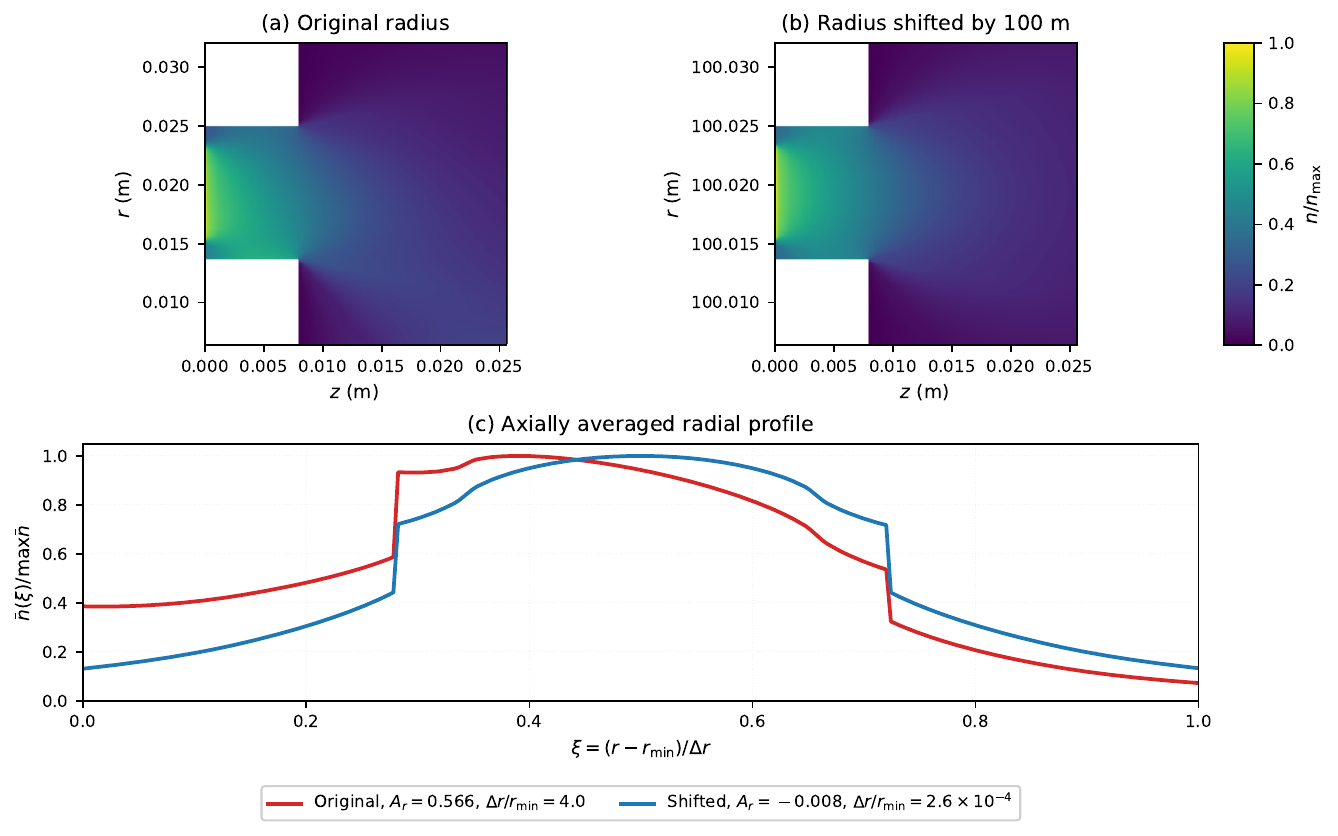}
    \caption{Radius-shift diagnostic. Panels (a,b) show normalized density fields;
    panel (c) compares global radial profiles and reports \(A_r\).}
    \label{fig:cylindrical_geometry_effect}
\end{figure}

To quantify the directional radial bias, the active computational domain is divided at
\(r_c=(r_{\min}+r_{\max})/2\) into inner- and outer-radius halves. A signed radial
asymmetry index is defined as
\begin{equation}
A_r=
\frac{
\langle n\rangle_{\mathrm{in}}-\langle n\rangle_{\mathrm{out}}
}{
\left(\langle n\rangle_{\mathrm{in}}+\langle n\rangle_{\mathrm{out}}\right)/2
}.
\label{eq:radial_asymmetry_index}
\end{equation}
where \(\langle n\rangle_{\mathrm{in}}\) and \(\langle n\rangle_{\mathrm{out}}\) are the
volume-averaged densities in the two halves of the full computational domain. A positive
value indicates that the density is biased toward the lower-radius side.

\Cref{fig:cylindrical_geometry_effect} shows that the original-radius case has a strong
lower-radius density bias, with \(A_r\approx0.566\). After shifting the same radial
thickness to a much larger radius, the bias is nearly removed, with
\(A_r\approx-0.008\). The radial profile also becomes much flatter in the logical
coordinate. This reduction indicates that a significant part of the lower-radius
enhancement in the baseline density maps is caused by the cylindrical area factor rather
than by a numerical symmetry error. Residual spatial variation is still expected because
the full neutral field also contains inlet-footprint, wall-reflection, and open-boundary
effects.

\FloatBarrier

\subsubsection{Comparison}
\label{subsubsec:comparison}

The comparison focuses on two related questions. First, the FM and $S_N$ preprocessors
are compared at the field level to assess whether they produce consistent neutral
transport fields under the same physical model. Second, for each preprocessor, the
density field obtained directly from the free-molecular calculation is compared with the
density obtained after solving the face-flux-closed continuity equation. The
corresponding within-method relative density recovery error is evaluated as
\begin{equation}
    \delta n
    =
    \frac{
    n^{\mathrm{cont}}
    -
    n^{\mathrm{ref}}
    }{
    \max
    \left(
    n^{\mathrm{ref}},
    n_{\mathrm{floor}}
    \right)
    },
    \label{eq:results_density_difference}
\end{equation}
where $n^{\mathrm{ref}}$ is the preprocessed reference density, $n^{\mathrm{cont}}$ is
the continuity-solved density, and $n_{\mathrm{floor}}$ is a small density floor used
only to avoid artificial amplification of relative errors in very low-density cells.

\begin{figure}[!tbp]
    \centering
    \includegraphics[width=0.78\linewidth]{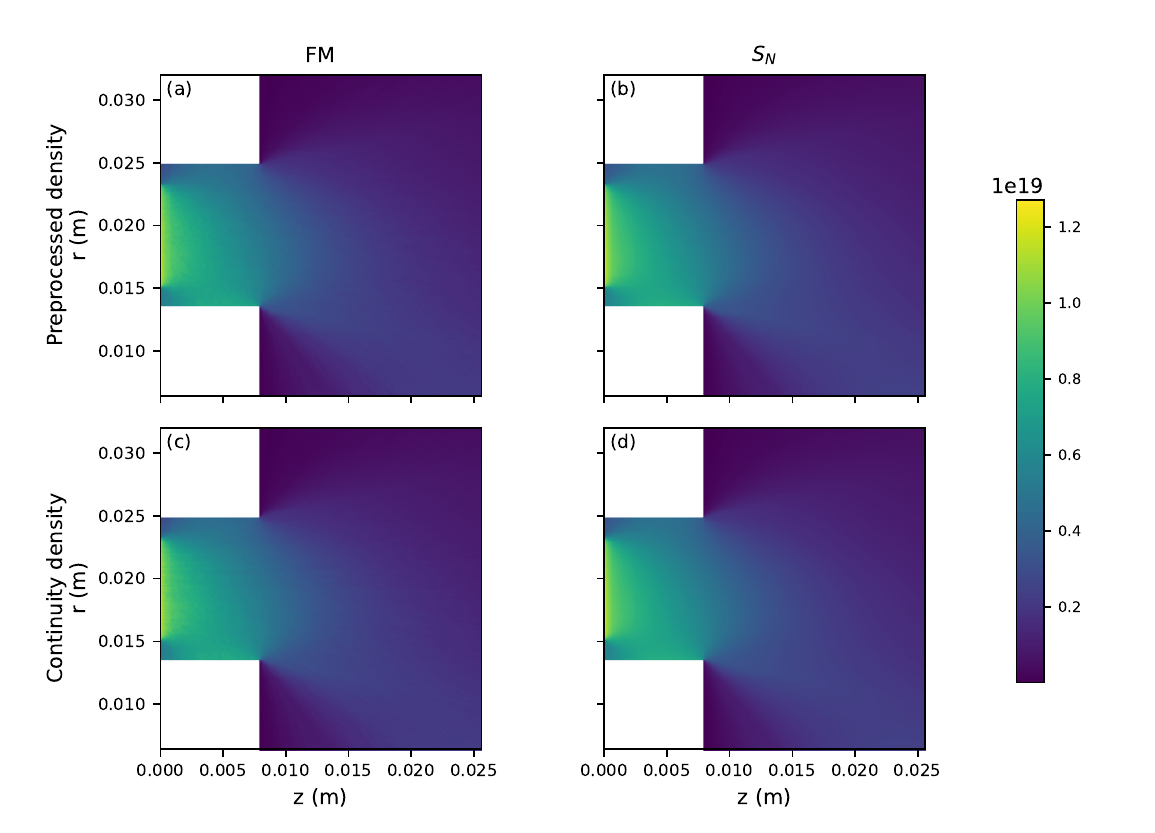}
    \caption{Preprocessed and continuity-solved density fields for the FM and
    $S_N$ preprocessors.}
    \label{fig:results_density_recovery_fields}
\end{figure}

\begin{figure}[!tbp]
    \centering
    \includegraphics[width=0.78\linewidth]{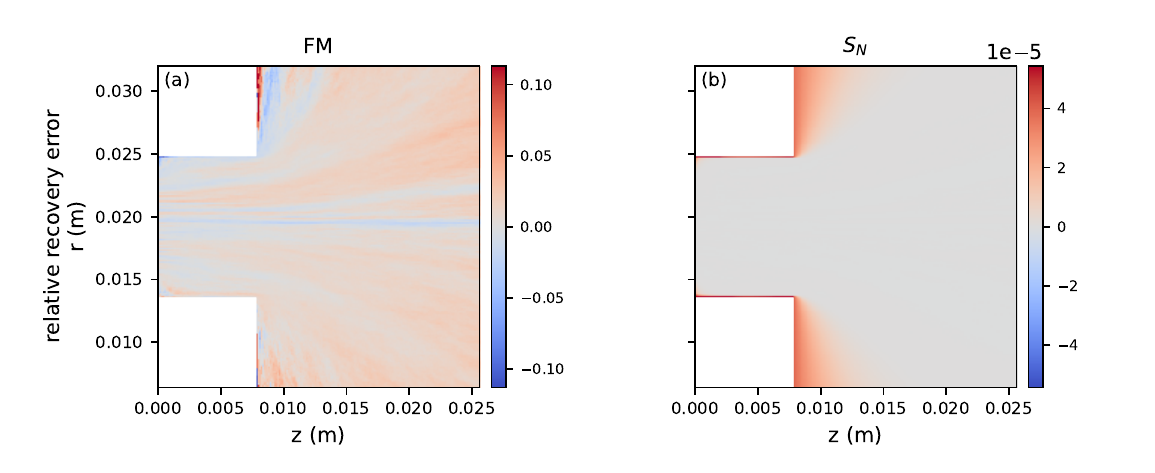}
    \caption{Density recovery errors for the FM and $S_N$ closures.}
    \label{fig:results_density_recovery_errors}
\end{figure}

The main field comparison is organized into two companion figures.
\Cref{fig:results_density_recovery_fields} compares the preprocessed and
continuity-solved density fields for the FM and $S_N$ preprocessors, using a common
density color scale across all four panels.
\Cref{fig:results_density_recovery_errors} shows the corresponding
within-method recovery error defined in Eq.~\eqref{eq:results_density_difference}. Each
panel uses its own adjacent color scale because the FM and $S_N$ recovery errors have
different characteristic magnitudes.

\Cref{fig:results_density_recovery_fields} shows that the two preprocessors
produce similar overall density structures. In both cases, the neutral density is
largest near the inlet band and decreases downstream as particles leave through the open
boundaries. Both methods also show a higher density toward the lower-radius side. As
shown in \cref{subsubsec:cylindrical_geometry_effect}, this trend is consistent with
cylindrical area weighting and does not by itself indicate a loss of geometric
consistency. For each method, the continuity-solved field closely matches its
corresponding preprocessed field, indicating that the face-flux handoff preserves the
dominant free-molecular transport pattern.

\Cref{fig:results_density_recovery_errors} quantifies the remaining recovery
error. The FM recovery error is visibly larger and contains small-scale fluctuations
associated with the particle-based face-flux statistics. In contrast, the $S_N$ recovery
error is several orders of magnitude smaller over most of the domain, reflecting the
deterministic consistency between the reconstructed face fluxes and the continuity
closure.

\begin{figure}[!hbp]
    \centering
    \includegraphics[width=0.94\linewidth]{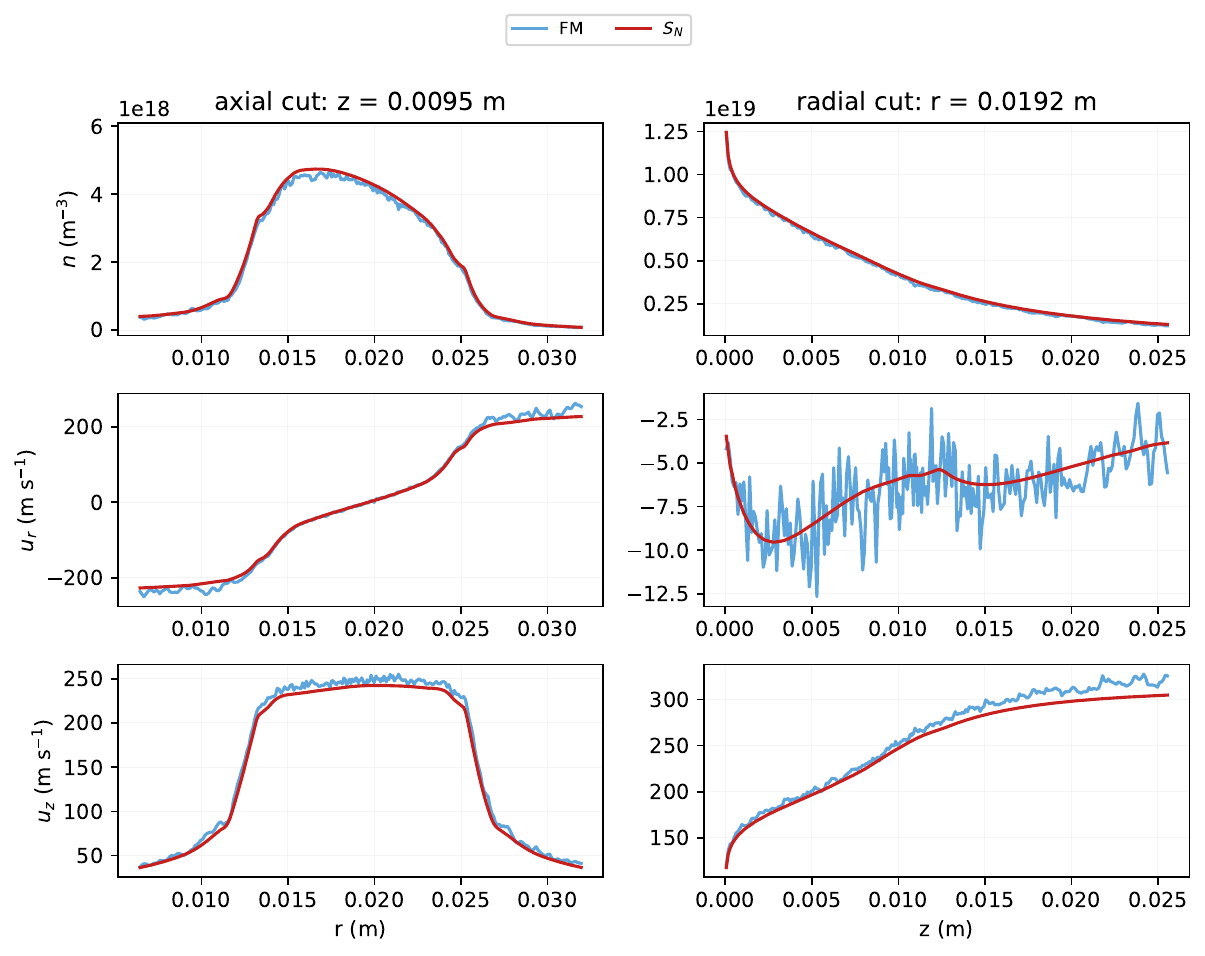}
    \caption{Density and velocity profiles along one axial cut and one radial
    cut.}
    \label{fig:results_radial_profiles}
\end{figure}

Additional one-dimensional profiles are extracted along one axial cut and one radial
cut. The compared quantities are the neutral density and the radial and axial mean
velocities reconstructed from each preprocessor. These profiles show the moment fields
implied by the FM and $S_N$ preprocessing strategies without mixing them with the
separate face-flux handoff diagnostics.
\Cref{fig:results_radial_profiles} compares these cuts. The $S_N$ density
profile remains slightly above the FM one, consistent with the baseline inventory ratio
of about $1.04$ at the selected working resolution. The velocity profiles show the same
overall trend, while the radial-velocity cut is visually noisier on the FM side because
$u_r$ is a smaller moment and is therefore more sensitive to Monte Carlo sampling noise.

The quantitative recovery and conservation metrics are summarized using density recovery
errors and global flux-balance errors. The density recovery metrics compare the
continuity-solved density with the preprocessed reference density. The volume-weighted
$L_1(n)$ and $L_2(n)$ errors measure the mean absolute and root-mean-square recovery
errors, respectively, while $L_\infty(n)$ reports the maximum pointwise relative error
using the same density floor as Eq.~\eqref{eq:results_density_difference}. Because
$L_\infty(n)$ is sensitive to isolated cells, it is used only as a worst-case indicator;
$L_1(n)$ and $L_2(n)$ are used to assess the overall recovery behavior. In the
no-ionization baseline case, the global flux balance is measured by comparing the
imposed inlet flux with the total outgoing open-boundary flux,
\begin{equation}
    \epsilon_{\Phi}
    =
    \frac{
    \left|
    \Phi_{\mathrm{in}}
    -
    \Phi_{\mathrm{out}}
    \right|
    }{
    \Phi_{\mathrm{in}}
    } .
    \label{eq:results_global_balance}
\end{equation}
Together, the density recovery errors and global balance error provide the quantitative
metrics for the baseline comparison, as summarized in
\cref{tab:baseline_metrics}.

\begin{table}[!tbp]
    \centering
    \caption{Baseline density-recovery and face-flux consistency metrics.}
    \label{tab:baseline_metrics}
    \small
    \begin{tabular}{lcccc}
\toprule
Method & $L_1(n)$ & $L_2(n)$ & $L_\infty(n)$ & $\epsilon_\Phi$ \\
\midrule
FM & 8.75e-03 & 9.31e-03 & 3.57e-01 & 1.44e-02 \\
$S_N$ & 7.23e-07 & 5.88e-06 & 1.26e-04 & 6.64e-07 \\
\bottomrule
\end{tabular}

\end{table}

Overall, the baseline comparison indicates that the $S_N$ preprocessor captures the
dominant free-molecular density and velocity structures obtained from the FM reference.
The continuity solve recovers the corresponding preprocessed density for each method,
confirming that the face-flux handoff is consistent at the reduced-continuity level. The
remaining differences between the two preprocessed fields appear mainly as a small
systematic inventory difference and localized particle-sampling fluctuations in the FM
moments.

\subsection{Parametric Study}
\label{subsec:parametric_study}

The parametric study is organized in two steps. First, the numerical resolution of the
deterministic $S_N$ preprocessor is examined, since the angular and speed quadrature
determine the accuracy of the deterministic reference field. The remaining scans then
vary physical or modeling parameters using the selected baseline resolution.

\subsubsection{Angular and Speed Resolution}
\label{subsubsec:param_resolution}

\begin{figure}[!tbp]
    \centering
    \includegraphics[width=0.92\linewidth]{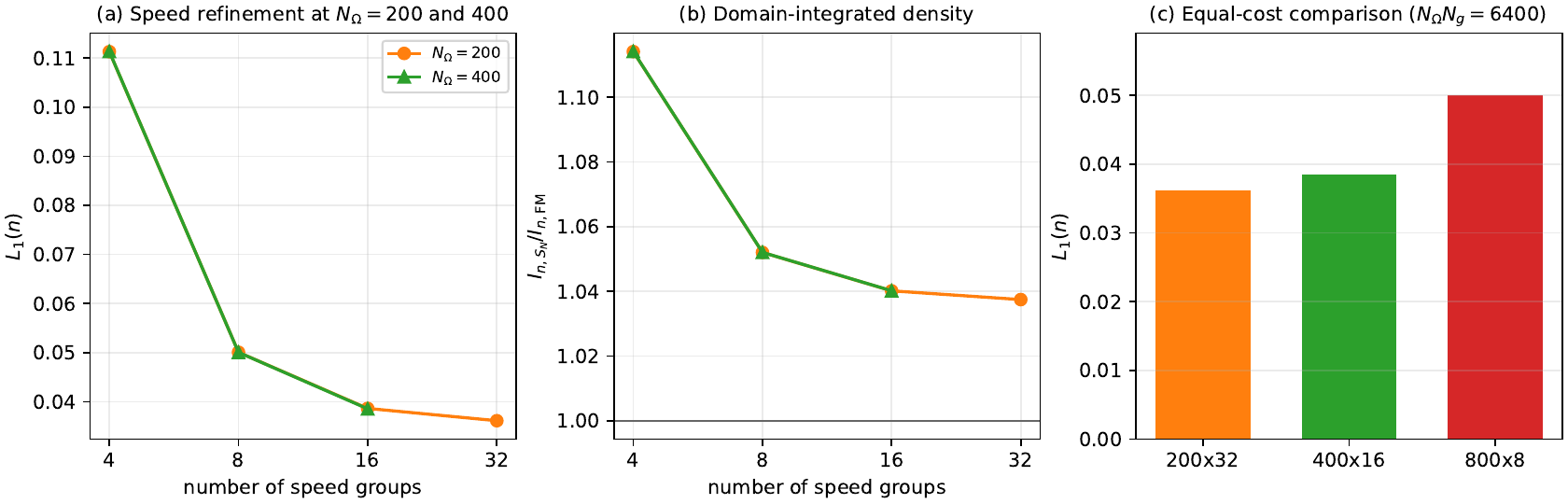}
    \caption{Velocity-space resolution sensitivity for the baseline case
    ($\alpha_d=0.7$).}
    \label{fig:sn_resolution_alpha07_summary}
\end{figure}

The velocity-space resolution of the deterministic $S_N$ preprocessor is controlled by
the number of angular ordinates $N_\Omega$ and the number of speed groups $N_g$. For the
baseline mixed-reflection wall condition $\alpha_d=0.7$, the resolution sensitivity is
evaluated by comparing the deterministic density field with the particle-based FM
face-flux reference while varying the angular-speed quadrature.
\Cref{fig:sn_resolution_alpha07_summary} and \cref{tab:sn_resolution_alpha07} summarize
the resulting density errors and inventory ratios.

For this mixed-reflection baseline, the density error is more sensitive to the
speed-group resolution than to the angular resolution. This trend is observed
consistently in both the $N_\Omega=200$ and $N_\Omega=400$ series. At $N_\Omega=400$,
increasing $N_g$ from $4$ to $8$ reduces the mainstream $L_1(n)$ density error from
about $0.11$ to $0.05$, and increasing $N_g$ from $8$ to $16$ further reduces it to
about $0.04$. The integrated neutral inventory follows the same trend, with the
overprediction decreasing from about $11\%$ at $N_g=4$ to about $4\%$ at $N_g=16$. The
$N_\Omega=200$ series shows essentially the same behavior and extends the
speed-refinement trend to $N_g=32$, for which the remaining improvement becomes small.
This behavior indicates that, once a sufficient angular resolution is used, the
residence-time-weighted density field in the present baseline configuration is mainly
controlled by the speed quadrature through the different residence-time weights of slow
and fast neutrals.

The equal-cost comparison among $200\times32$, $400\times16$, and $800\times8$ then
clarifies how a fixed budget of discrete velocity states should be allocated. All three
cases use $N_\Omega N_g=6400$, but their errors are not the same. The $200\times32$ case
gives the smallest mainstream density error, showing that additional speed resolution
can still provide a slight improvement. However, the improvement over $400\times16$ is
marginal: the mainstream $L_1(n)$ error decreases only from about $0.0405$ to $0.0380$,
and the total neutral inventory ratio decreases only from about $1.040$ to $1.037$. Here
the inventory ratio denotes the domain-integrated neutral density $I_{n,S_N}=\int_\Omega
n_{S_N}\,dV$ normalized by the FM reference value $I_{n,\mathrm{FM}}=\int_\Omega
n_{\mathrm{FM}}\,dV$. In contrast, $800\times8$ gives a larger density error because the
reduced speed resolution increases the residence-time-weighting error. Therefore,
increasing angular resolution at the expense of speed resolution is less effective for
the present baseline than keeping a larger number of speed groups.

Based on these observations, $N_\Omega=400$ and $N_g=16$ are selected as the baseline
working resolution for the remaining parameter scans. This choice does not represent a
claim of strict full convergence in speed space. Rather, it provides a practical
near-plateau speed resolution while retaining a higher angular resolution than
$200\times32$. This is desirable for the subsequent wall-reflection scans, because lower
diffuse fractions are expected to be more sensitive to angular discretization and
directional coherence. The $N_g=32$ result is therefore used as a supplementary
convergence check rather than as the default working resolution.

\begin{table}[!tbp]
    \centering
    \caption{Representative resolution metrics for the baseline
    mixed-reflection case.}
    \label{tab:sn_resolution_alpha07}
    \small
    \begin{tabular}{cccccc}
        \toprule
        Case & $N_\Omega$ & $N_g$ & $N_\Omega N_g$ & $L_1(n)$ & $I_{n,S_N}/I_{n,\mathrm{FM}}$ \\
        \midrule
        $200\times4$ & 200 & 4  & 800  & 0.111 & 1.114 \\
        $200\times8$ & 200 & 8  & 1600 & 0.050 & 1.052 \\
        $200\times16$ & 200 & 16 & 3200 & 0.0406 & 1.0402 \\
        $200\times32$ & 200 & 32 & 6400 & 0.0380 & 1.0375 \\
        $400\times4$ & 400 & 4  & 1600 & 0.111 & 1.114 \\
        $400\times8$ & 400 & 8  & 3200 & 0.050 & 1.052 \\
        $\mathbf{400\times16}$ & \textbf{400} & \textbf{16} & \textbf{6400} & \textbf{0.0405} & \textbf{1.0402} \\
        $800\times8$ & 800 & 8  & 6400 & 0.0500 & 1.0521 \\
        \bottomrule
    \end{tabular}
\end{table}

\subsubsection{Diffuse Reflection Fraction}
\label{subsubsec:param_diffuse_fraction}

The diffuse reflection fraction $\alpha_d$ controls the degree to which neutral atoms
lose directional memory at the wall. When $\alpha_d=0$, the wall reflection is purely
specular and the outgoing angular information is preserved through mirror reflection.
When $\alpha_d=1$, the reflected particles are fully accommodated at the wall and
re-emitted according to the wall-temperature diffuse distribution. Intermediate values
represent mixed reflection, in which a fraction of the outgoing wall flux is
redistributed over the incoming half-space. Varying $\alpha_d$ therefore provides a
direct way to examine how wall accommodation modifies both the physical neutral
transport pattern and the numerical behavior of the deterministic angular
discretization. This scan is used to clarify how wall accommodation controls directional
coherence and the severity of the $S_N$ ray effect.

\begin{figure}[!tbp]
    \centering
    \includegraphics[width=0.92\linewidth]{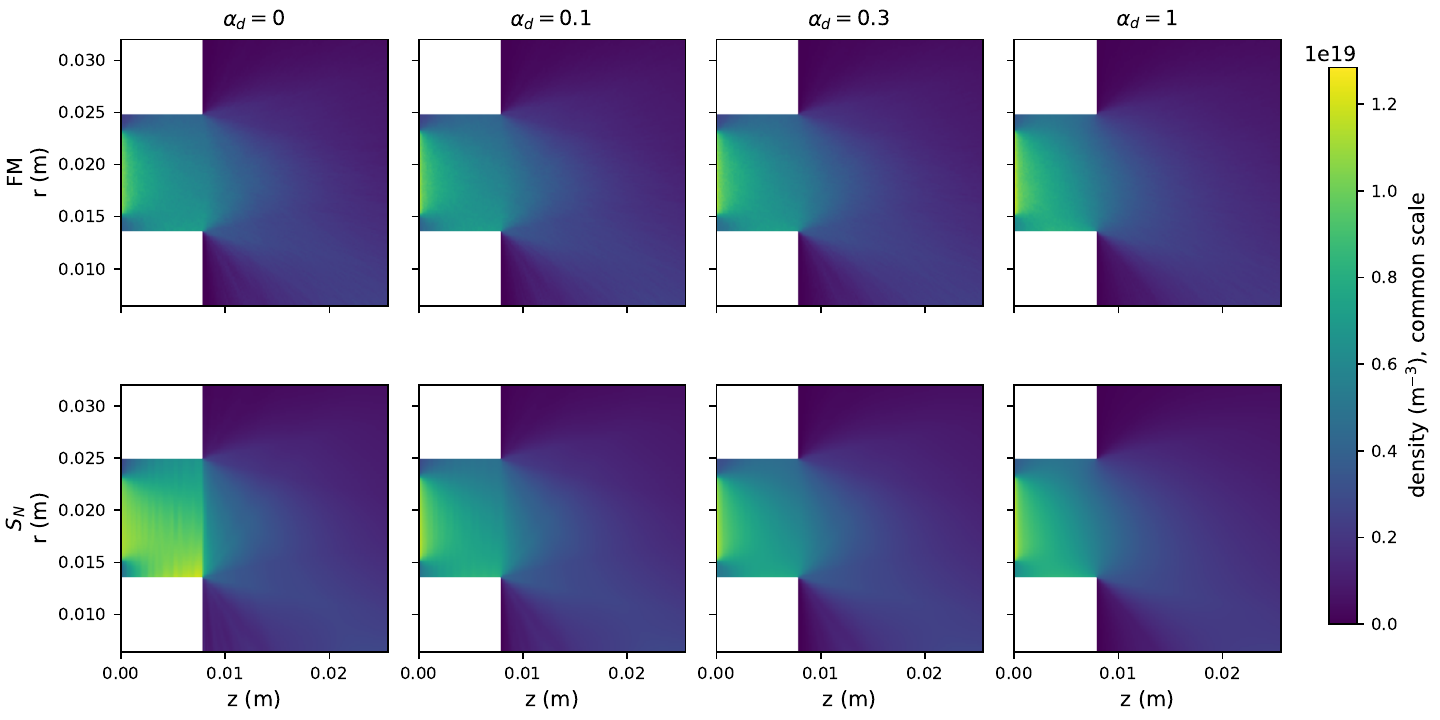}
    \caption{Density fields for selected diffuse reflection fractions.}
    \label{fig:diffuse_reflection_density}
\end{figure}

\Cref{fig:diffuse_reflection_density} shows representative density fields for
$\alpha_d=0$, $0.1$, $0.3$, and $1.0$. In the pure-specular case, the neutral field
retains stronger directional coherence. The density distribution is more structured, and
ray-like features are more visible because the wall reflection does not significantly
redistribute the angular flux, consistent with the classical ray effect of
discrete-ordinates methods \cite{AdamsFastiterativeme2002,CamminadyRayeffectmitiga2019}.
As $\alpha_d$ increases, the wall provides stronger angular re-emission and the density
field becomes smoother. Physically, this reflects the gradual loss of directional memory
at the wall: outgoing particles are progressively re-injected into a broader set of
incoming directions rather than following a nearly deterministic mirror path.

\begin{figure}[!tbp]
    \centering
    \includegraphics[width=0.84\linewidth]{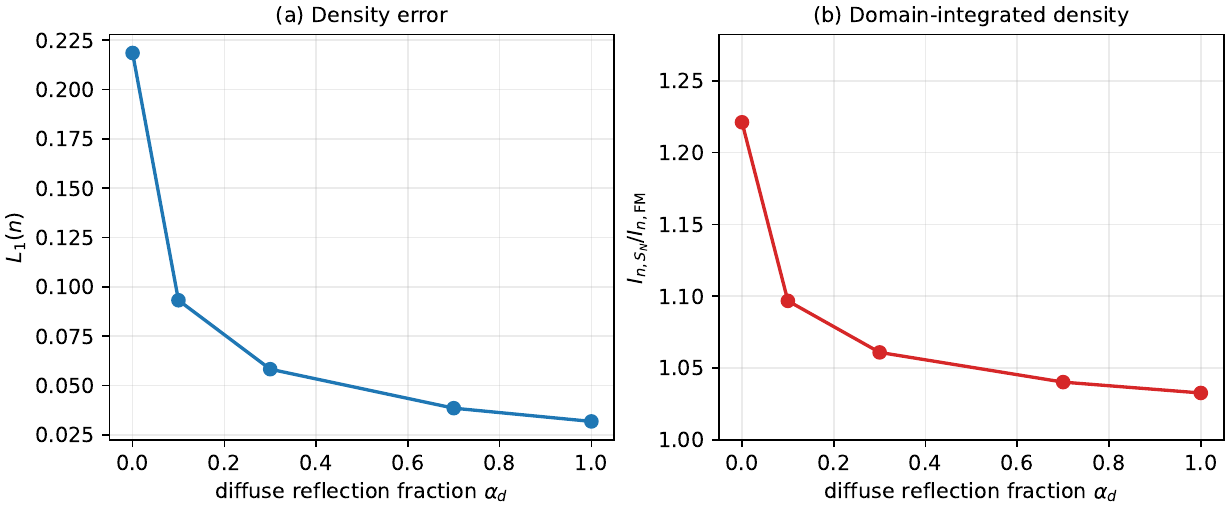}
    \caption{Mainstream density error and total neutral inventory ratio versus
    diffuse reflection fraction.}
    \label{fig:diffuse_reflection_metrics}
\end{figure}

The quantitative comparison in \cref{fig:diffuse_reflection_metrics} and
\cref{tab:diffuse_reflection_metrics} shows the same transition over the full
scan. The mainstream density error is largest in the pure-specular limit, with
$L_1(n)\approx 0.219$. Introducing a modest diffuse component produces the largest
improvement: by $\alpha_d=0.1$, the error decreases to about $0.093$ and the total
neutral inventory ratio decreases from about $1.221$ to about $1.097$. This indicates
that the deterministic solution changes from a strongly directionally coherent regime to
a much more stable mixed-reflection regime once wall re-emission begins to redistribute
the outgoing angular flux. Further increasing $\alpha_d$ continues to improve the
agreement, but the improvement becomes more gradual: the error decreases to about
$0.058$ at $\alpha_d=0.3$ and $0.032$ in the fully diffuse case. The intermediate points
reported in \cref{tab:diffuse_reflection_metrics} confirm that this change is monotonic.

\begin{table}[!tbp]
    \centering
    \caption{Representative metrics for the diffuse-reflection scan.}
    \label{tab:diffuse_reflection_metrics}
    \small
    \begin{tabular}{ccccc}
        \toprule
        $\alpha_d$ & $L_1(n)$ & $L_2(n)$ & $I_{n,S_N}/I_{n,\mathrm{FM}}$ & Regime \\
        \midrule
        0    & 0.219 & 0.344 & 1.221 & pure specular \\
        0.1  & 0.093 & 0.112 & 1.097 & transition \\
        0.3  & 0.058 & 0.059 & 1.061 & mixed \\
        0.7  & 0.039 & 0.036 & 1.040 & baseline mixed \\
        1.0  & 0.032 & 0.029 & 1.033 & diffuse \\
        \bottomrule
    \end{tabular}
\end{table}

The pure-specular case is also more sensitive to angular resolution. Since no diffuse
re-emission is present, discrete ordinate directions remain coherent after repeated wall
reflections. A separate angular-resolution test at $\alpha_d=0$ confirms that increasing
$N_\Omega$ reduces the visibility of these ray-like structures, while speed refinement
mainly affects the density magnitude through residence-time weighting. Once a finite
diffuse fraction is introduced, the wall re-emission redistributes the outgoing flux
over the incoming half-space and naturally suppresses this coherence. For this reason,
the pure-specular and near-specular cases are interpreted here as limiting sensitivity
tests, whereas the mixed-reflection cases are more representative of the intended wall
model. The baseline choice belongs to this stable mixed-reflection regime and is
therefore used for the main comparison and the subsequent parameter scans.

\subsubsection{Inlet Density and Axial Drift}
\label{subsubsec:param_inlet_flux}

The inlet-density scan is performed here as a combined variation of inlet density and
axial drift velocity. The inlet width, inlet temperature, wall temperature,
wall-reflection coefficient, and deterministic resolution are kept fixed, while the
inlet density is varied over $n_{\mathrm{in}}=1.0\times10^{18}$, $5.0\times10^{18}$, and
$1.0\times10^{19}\ \mathrm{m^{-3}}$, and the axial drift velocity is varied over
$u_{z0}=300$, $600$, and $900\ \mathrm{m\,s^{-1}}$. The baseline setting therefore
corresponds to the middle density level and the lowest-drift branch of this $3\times3$
scan. The density branch tests the expected linear scaling of collisionless
free-molecular transport, while the drift branch tests the residence-time reduction and
downstream shift caused by stronger inlet axial momentum.

\begin{figure}[!tbp]
    \centering
    \includegraphics[width=0.82\linewidth]{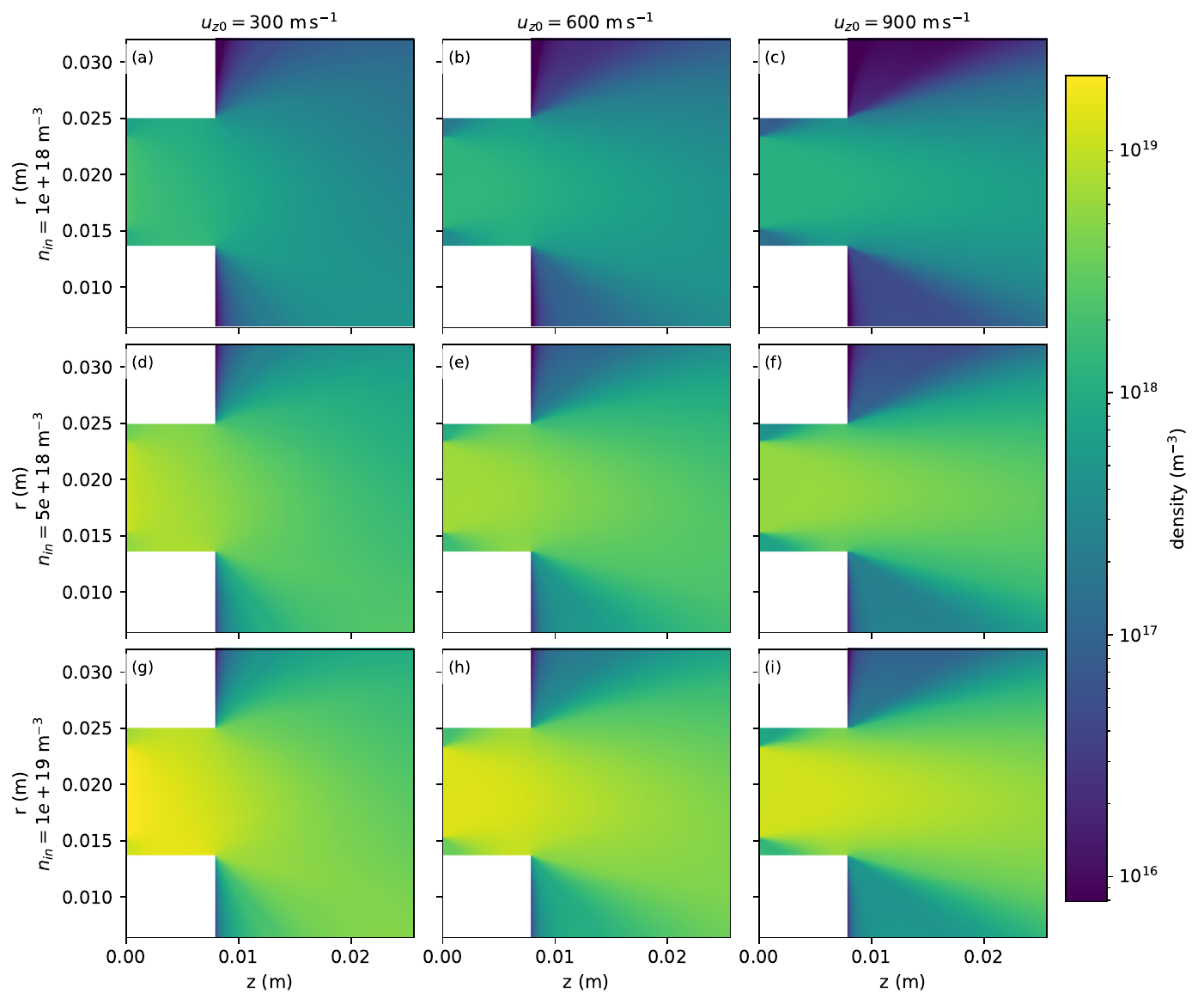}
    \caption{Deterministic density fields for the $3\times3$ inlet-density and
    axial-drift scan. A common logarithmic color scale is used.}
    \label{fig:inlet_flux_drift_density_fields}
\end{figure}

\Cref{fig:inlet_flux_drift_density_fields} shows the full $3\times3$ matrix
of deterministic density fields. The rows correspond to
$n_{\mathrm{in}}=1.0\times10^{18}$, $5.0\times10^{18}$, and $1.0\times10^{19}\
\mathrm{m^{-3}}$, and the columns correspond to $u_{z0}=300$, $600$, and $900\
\mathrm{m\,s^{-1}}$. Reading down each column, the dominant trend with inlet density is
nearly linear scaling of the density-field amplitude, which is expected in the
collisionless free-molecular limit when the wall model is fixed. At a fixed drift
velocity, increasing $n_{\mathrm{in}}$ by a factor of ten increases both the peak
density and the total neutral inventory by approximately the same factor. Reading across
each row, increasing the axial drift shifts the distribution downstream and reduces the
degree of near-anode accumulation.

\begin{figure}[!tbp]
    \centering
    \includegraphics[width=0.82\linewidth]{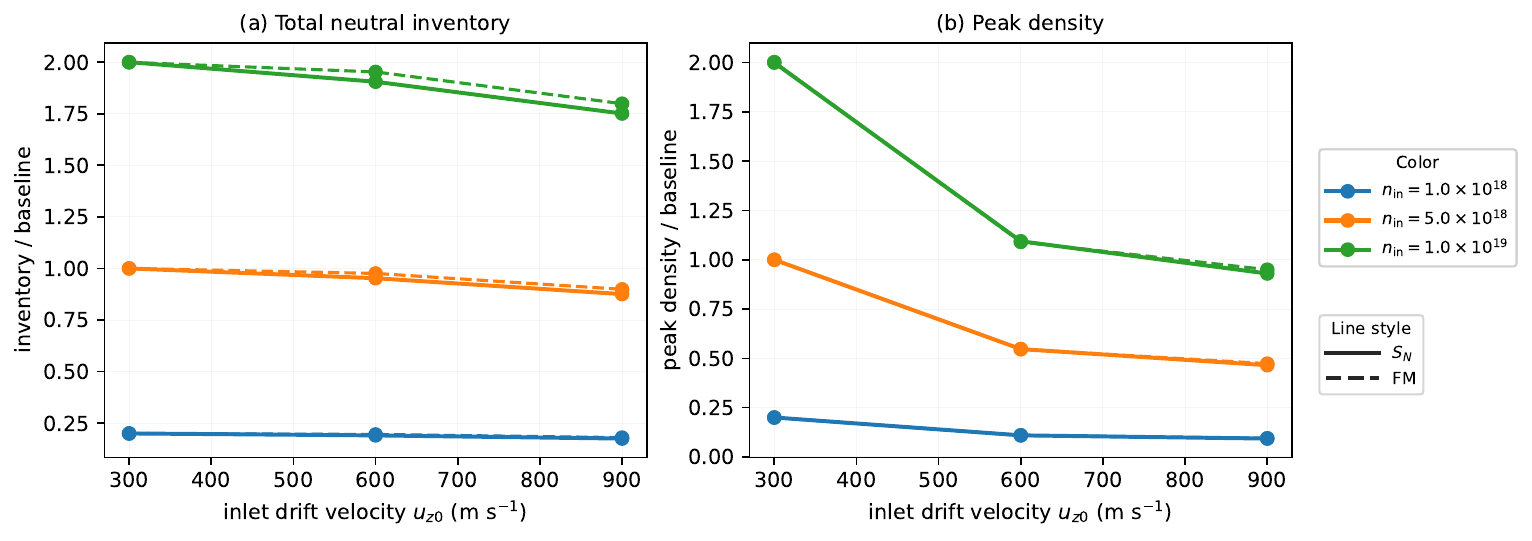}
    \caption{Normalized inventory and peak density for the inlet-density and
    axial-drift scan.}
    \label{fig:inlet_flux_drift_metrics}
\end{figure}

The axial drift velocity affects both the injected axial momentum and the residence time
of neutrals in the computational domain. In the present scan, the residence-time
reduction dominates the density response: as $u_{z0}$ is increased, the density-weighted
mean axial velocity increases, but the total neutral inventory and peak density both
decrease for a fixed inlet density.
\Cref{fig:inlet_flux_drift_metrics} shows that this trend is systematic across
the three inlet-density levels. Solid lines show the $S_N$ results, and dashed lines
show the FM reference results. Each method is normalized by its own baseline value, so
the figure compares the relative response to inlet drift rather than the absolute
$S_N$-to-FM inventory ratio. For example, at fixed $n_{\mathrm{in}}=5.0\times10^{18}\
\mathrm{m^{-3}}$, the total inventory decreases from the baseline value to about $95\%$
at $u_{z0}=600\ \mathrm{m\,s^{-1}}$ and to about $87.5\%$ at $u_{z0}=900\
\mathrm{m\,s^{-1}}$. The peak density decreases even more strongly, which indicates that
larger axial drift not only shortens the mean residence time but also reduces the degree
of near-anode accumulation.

Taken together, these results show that inlet density and axial drift affect the neutral
field in qualitatively different ways. The inlet density primarily sets the overall
throughput and produces an approximately linear amplitude scaling, whereas the axial
drift velocity controls how rapidly neutrals are transported downstream and therefore
how strongly they accumulate in the near-anode region. This combined scan is therefore
best interpreted as a joint throughput-and-directionality sensitivity study rather than
as a pure inlet flux-amplitude perturbation.

\subsubsection{Inlet Width}
\label{subsubsec:param_inlet_width}

The inlet-width scan changes the radial extent of the imposed inlet band while keeping
the unit-area inlet model fixed. The inlet density, inlet temperature, wall temperature,
axial drift velocity, wall-reflection coefficient, and deterministic resolution are
unchanged. Therefore, increasing the inlet width changes both the spatial footprint of
the source and the total injected neutral flux. This scan should therefore be
interpreted as a combined footprint-and-throughput sensitivity study rather than as a
purely geometric perturbation. It is used to separate the source-footprint effect from
the leading throughput scaling as far as possible under fixed unit-area inlet conditions.

\Cref{tab:inlet_width_cases} lists the three inlet-width cases. The baseline
case W2 is normalized to unity. The narrow case W1 has about $50.7\%$ of the baseline
inlet area and therefore about $50.7\%$ of the baseline total inlet flux. The wide case
W3 has about $140.9\%$ of the baseline inlet area and therefore about $140.9\%$ of the
baseline total inlet flux.

\begin{table}[!tbp]
    \centering
    \caption{Inlet-width scan with fixed unit-area inlet conditions.}
    \label{tab:inlet_width_cases}
    \small
    \begin{tabularx}{0.72\linewidth}{@{}cYYY@{}}
        \toprule
        Case & Width (m) & Relative inlet area & Relative total inlet flux \\
        \midrule
        W1 & 0.0040664391 & 0.5069 & 0.5069 \\
        W2 & 0.0080367644 & 1.0000 & 1.0000 \\
        W3 & 0.0113108448 & 1.4094 & 1.4094 \\
        \bottomrule
    \end{tabularx}
\end{table}

The density-field comparison shows that the inlet-width variation changes both the
magnitude and the spatial extent of the near-anode neutral population. In the
narrow-inlet case, the source is more concentrated and the injected neutral population
remains more localized near the inlet footprint. As the inlet width increases, the
near-anode high-density region expands radially and the downstream density field becomes
broader. This reflects both the larger injected neutral inventory and the wider set of
free-molecular trajectories populated directly from the inlet.

\begin{figure}[!tbp]
    \centering
    \includegraphics[width=0.82\linewidth]{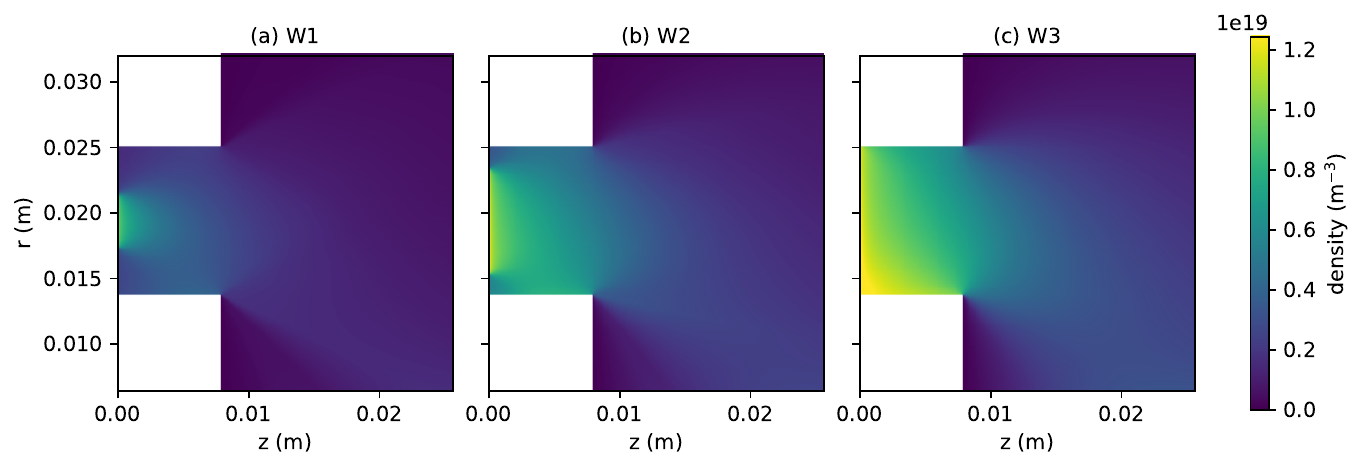}
    \caption{Density fields for different inlet widths. The unit-area inlet
    condition is kept fixed.}
    \label{fig:inlet_width_density_fields}
\end{figure}

\begin{figure}[!tbp]
    \centering
    \includegraphics[width=0.88\linewidth]{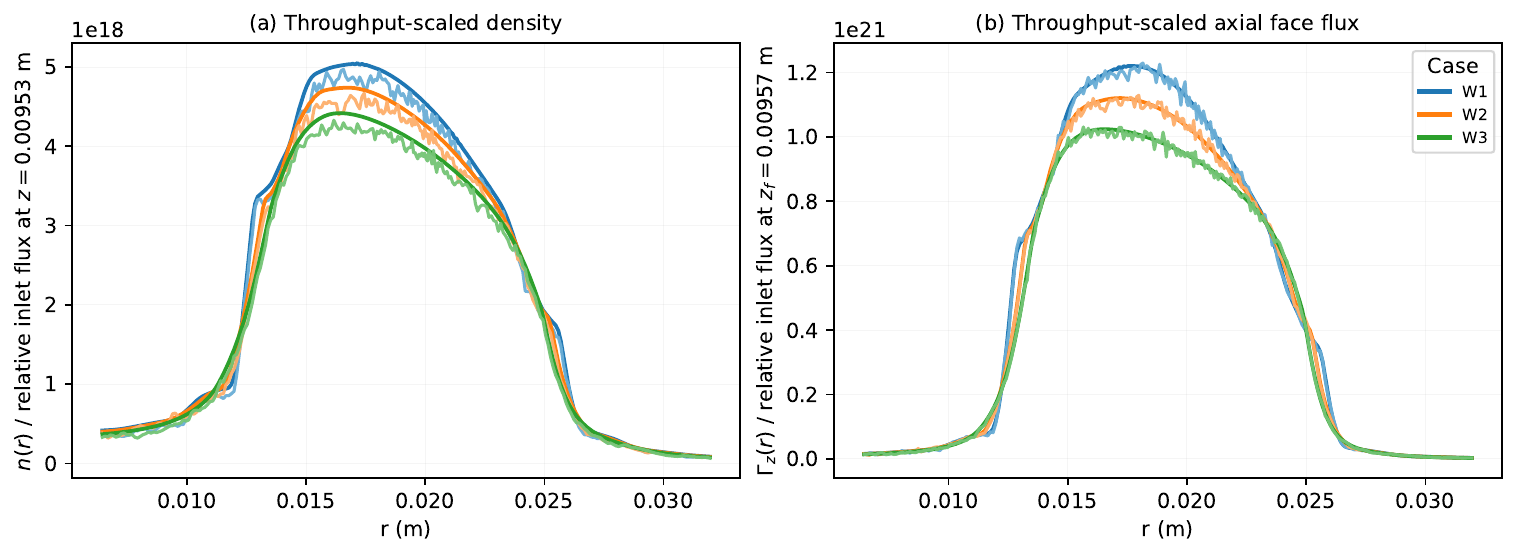}
    \caption{Density and axial face-flux profiles normalized by relative inlet
    flux.}
    \label{fig:inlet_width_profiles}
\end{figure}

The profile comparison in \cref{fig:inlet_width_profiles} removes the leading throughput
effect by dividing each curve by the relative inlet flux, defined here as the relative
total inlet flux listed in \cref{tab:inlet_width_cases}. Since this divisor is
dimensionless, the plotted quantities retain the units of density and face flux. The
density profile is extracted at $z\approx 9.53\times10^{-3}\ \mathrm{m}$, and the axial
face-normal flux is evaluated on the nearest selected internal axial face at $z_f\approx
9.57\times10^{-3}\ \mathrm{m}$. After this scaling, the wider inlet still produces a
broader radial footprint, while the narrow inlet remains more concentrated. This
indicates that inlet width changes the free-molecular trajectory population, not only
the total injected amount. The FM and $S_N$ profiles follow the same ordering across W1,
W2, and W3.

Because the unit-area inlet model is kept fixed in the present scan, the total injected
neutral flux also changes with inlet width. A fixed-total-flux inlet-width scan would
isolate the geometric footprint effect more cleanly and may be considered separately.
The present results should therefore be interpreted as the combined effect of source
footprint and inlet throughput.

\subsection{Ionization-Loss Extension}
\label{subsec:ionization_loss}

A prescribed ionization-loss case is introduced to examine how a volumetric neutral sink
modifies the preprocessed free-molecular transport field. The purpose of this test is
not to model a self-consistent plasma ionization process, but to provide a controlled
neutral-loss field that can be applied consistently to the particle-based and
deterministic preprocessors. The ionization frequency is prescribed as
\begin{equation}
    \nu_{\mathrm{ion}}(r,z)
    =
    \nu_{\max}\mathcal{I}(r,z),
    \label{eq:ionization_frequency_model}
\end{equation}
where $\mathcal{I}(r,z)$ is a localized cosine-window function. Inside the specified
ionization rectangle,
\begin{equation}
    \mathcal{I}(r,z)
    =
    \cos\left[
    \pi\frac{r-r_m}{r_2-r_1}
    \right]
    \cos\left[
    \pi\frac{z-z_m}{z_2-z_1}
    \right],
    \label{eq:ionization_window}
\end{equation}
and $\mathcal{I}=0$ outside this region. The center of the window is $r_m=(r_1+r_2)/2$
and $z_m=(z_1+z_2)/2$.

In the deterministic $S_N$ preprocessor, ionization is treated as an absorption term in
each speed-angle transport equation. For speed group $g$ and ordinate $a$, the steady
equation is written as
\begin{equation}
    v_g\OmegaVec_a\cdot\nabla_{r,z}\psi_{g,a}
    +
    \nu_{\mathrm{ion},i,k}\psi_{g,a}
    =
    0 .
    \label{eq:sn_ionization_transport}
\end{equation}
for a cell $K_{i,k}$ with prescribed cellwise loss frequency $\nu_{\mathrm{ion},i,k}$.
Equivalently, if the streaming operator is normalized by $v_g$, the absorption
coefficient appears as $\nu_{\mathrm{ion}}/v_g$. This form reflects the residence-time
effect: slower neutrals experience stronger attenuation over the same path length. In
the particle-based FM face-flux preprocessor, the same prescribed frequency is applied
as a stochastic neutral-removal process. For a particle trajectory segment of duration
$\Delta t_s$, the ionization probability is
\begin{equation}
    P_{\mathrm{ion},s}
    =
    1-\exp\left(-\nu_{\mathrm{ion},s}\Delta t_s\right),
    \label{eq:particle_ionization_probability}
\end{equation}
where $\nu_{\mathrm{ion},s}$ is evaluated along the segment. In the reduced continuity
equation, the corresponding neutral loss term is
\begin{equation}
    \frac{\partial n}{\partial t}
    +
    \nabla\cdot\boldsymbol{\Gamma}
    =
    -
    \nu_{\mathrm{ion}} n .
    \label{eq:ionization_continuity}
\end{equation}
Thus, the particle, deterministic, and reduced-continuity descriptions use the same
prescribed loss frequency, but apply it at different levels: individual particles,
angular-speed distribution functions, and density moments. In the continuity solve, the
loss term is treated semi-implicitly in pseudo-time, while the face-flux divergence is
evaluated from the preprocessed ionization-modified face fluxes.

\begin{figure}[!tbp]
    \centering
    \includegraphics[width=0.78\linewidth]{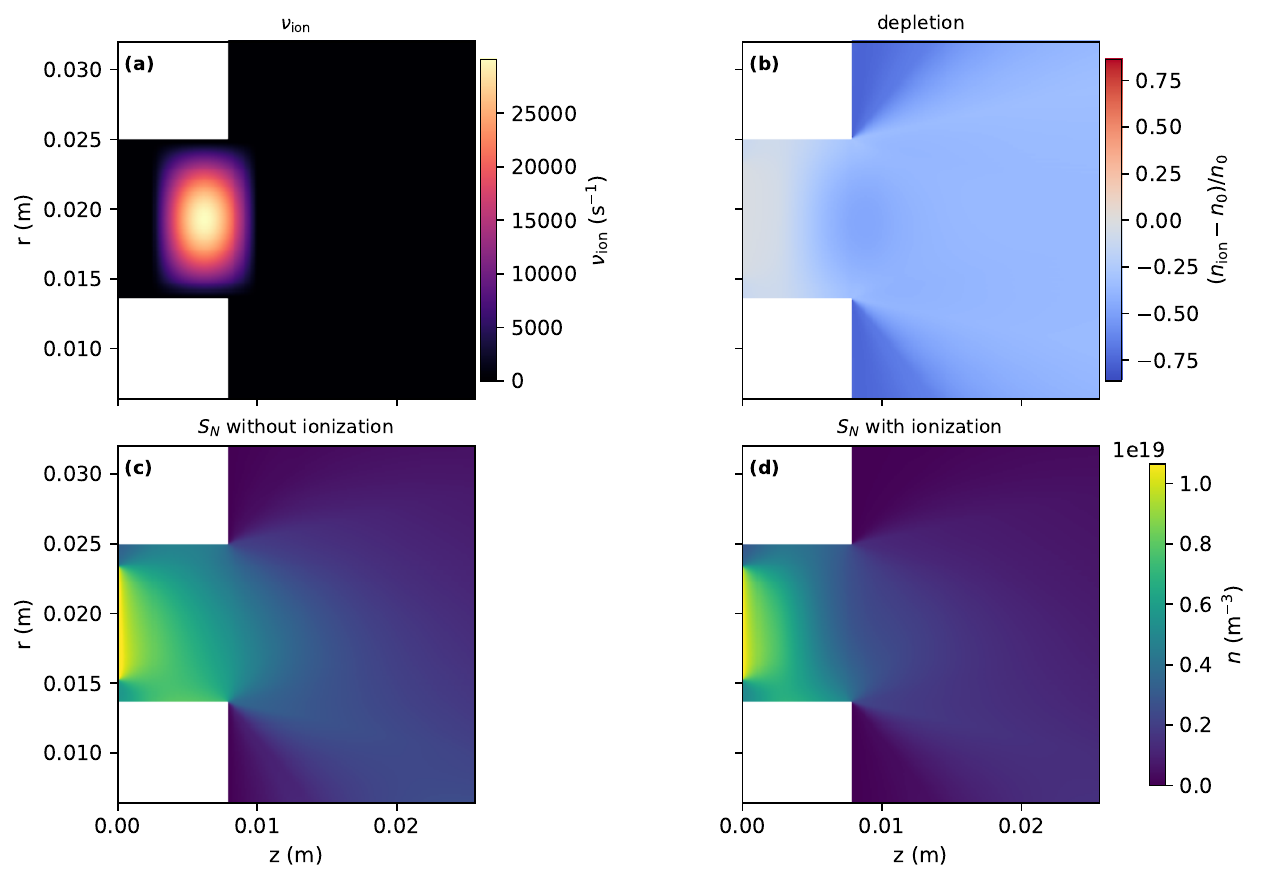}
    \caption{$S_N$ density response to the prescribed ionization-loss field.}
    \label{fig:ionization_density_response}
\end{figure}

The demonstration case uses $\nu_{\max}=3.0\times10^4\ \mathrm{s^{-1}}$. This value
represents a moderate prescribed neutral-loss case.
\Cref{fig:ionization_density_response} shows the prescribed ionization-frequency field
and the deterministic $S_N$ density response. The panels include the no-ionization $S_N$
density field, the $S_N$ density field with ionization, and the relative depletion
caused by the neutral loss. The ionization loss reduces the neutral density primarily
inside and downstream of the prescribed ionization region. The depletion pattern is not
confined to the loss region itself, because absorbed neutrals would otherwise contribute
to downstream free-molecular trajectories and wall-reflection paths.

\begin{table}[!tbp]
    \centering
    \caption{Ionization-loss window parameters.}
    \label{tab:ionization_loss_parameters}
    \small
    \begin{tabular}{ll}
        \toprule
        Parameter & Value \\
        \midrule
        $r_1$ (inner radial boundary) & $1.3683\times10^{-2}$ m \\
        $r_2$ (outer radial boundary) & $2.4672\times10^{-2}$ m \\
        $z_1$ (upstream axial boundary) & $2.4710\times10^{-3}$ m \\
        $z_2$ (downstream axial boundary) & $1.0022\times10^{-2}$ m \\
        $\nu_{\max}$ (peak ionization frequency) & $3.0\times10^{4}$ s$^{-1}$ \\
        \bottomrule
    \end{tabular}
\end{table}

\begin{figure}[!tbp]
    \centering
    \includegraphics[width=0.93\linewidth]{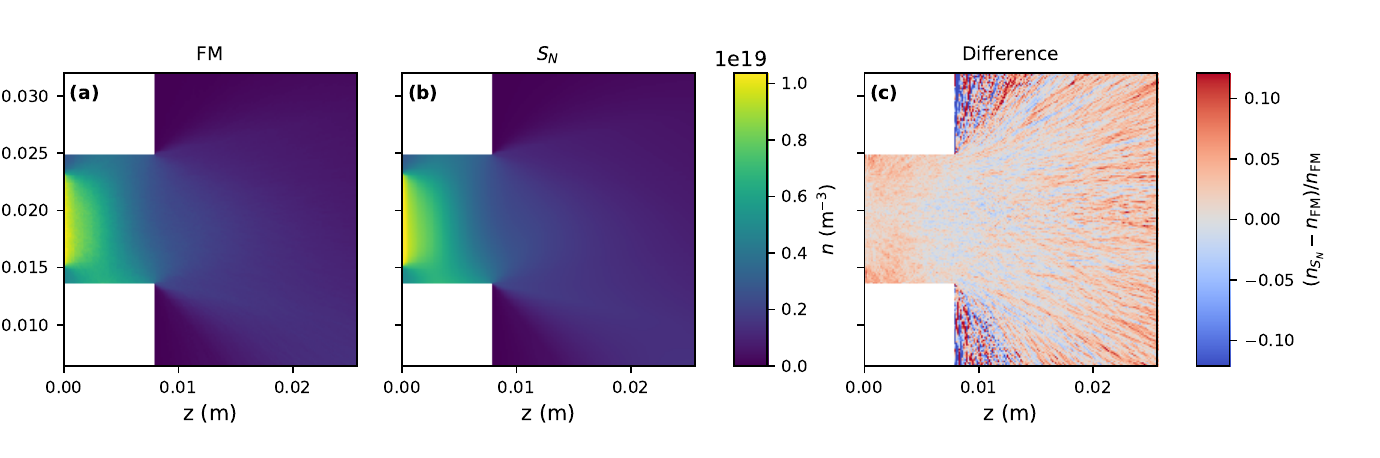}
    \caption{FM and deterministic $S_N$ density fields for the prescribed
    ionization-loss case.}
    \label{fig:ionization_faceflux_comparison}
\end{figure}

The ionization-modified deterministic result is also compared with the particle-based FM
face-flux preprocessor using the same prescribed $\nu_{\mathrm{ion}}$ field.
\Cref{fig:ionization_faceflux_comparison} shows the FM density, the deterministic $S_N$
density, and their relative difference. The two preprocessors give the same main density
structure for the present moderate-loss case. The remaining differences are mainly
localized near low-density edge regions and are small in the main-density region,
indicating that the prescribed neutral-loss field does not change the main consistency
trend observed in the no-ionization baseline comparison.

With ionization, the global neutral balance becomes
\begin{equation}
    \Phi_{\mathrm{in}}
    \approx
    \Phi_{\mathrm{out}}
    +
    \Phi_{\mathrm{ion}},
    \qquad
    \Phi_{\mathrm{ion}}
    =
    \int_{\mathcal{D}}
    \nu_{\mathrm{ion}}n\,\dd V .
    \label{eq:ionization_global_balance}
\end{equation}
For the present case, the integrated ionization loss is approximately $30.7\%$ of the
imposed inlet flux, and the open-boundary outflow plus volumetric ionization loss
satisfies the global balance to within the error reported in
\cref{tab:ionization_metrics}. This indicates that the prescribed volumetric loss is
represented consistently at the global-balance level.

\begin{figure}[!tbp]
    \centering
    \includegraphics[width=0.93\linewidth]{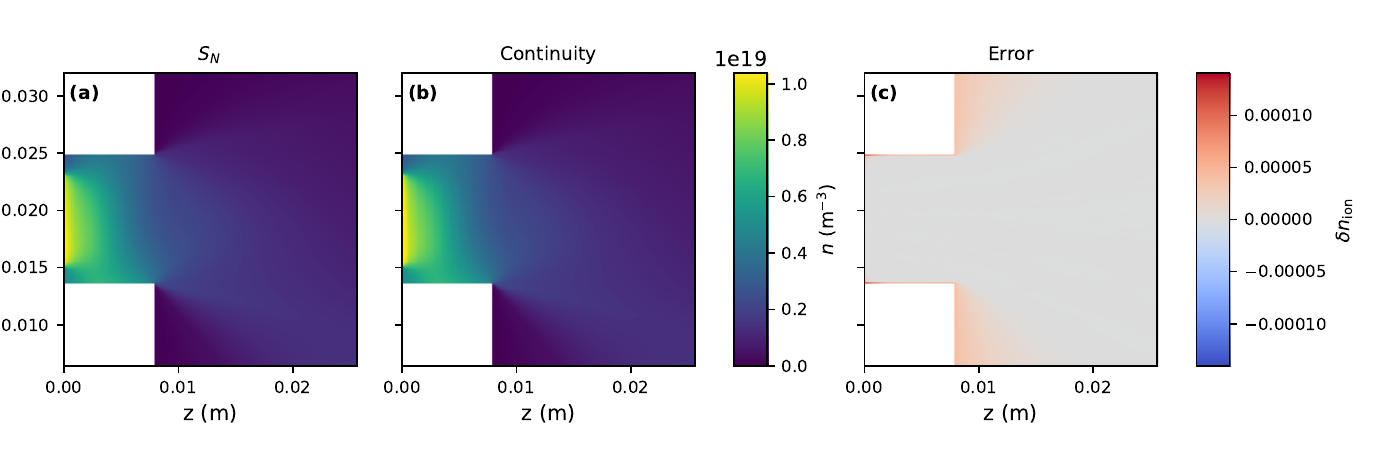}
    \caption{Continuity recovery for the prescribed ionization-loss case.}
    \label{fig:ionization_continuity_recovery}
\end{figure}

\Cref{fig:ionization_continuity_recovery} compares the ionization-modified
deterministic density field with the density obtained from the reduced continuity
equation. The continuity solve uses the same prescribed $\nu_{\mathrm{ion}}$ field and
the face fluxes produced by the deterministic preprocessor with ionization. The relative
recovery error is defined as
\begin{equation}
    \delta n_{\mathrm{ion}}
    =
    \frac{
    n_{\mathrm{cont,ion}}
    -
    n_{S_N,\mathrm{ion}}
    }{
    \max\left(n_{S_N,\mathrm{ion}}, n_{\mathrm{floor}}\right)
    } .
    \label{eq:ionization_recovery_error}
\end{equation}
where $n_{\mathrm{floor}}$ is the same small density floor used in the other recovery
comparisons. For $\nu_{\max}=3.0\times10^4\ \mathrm{s^{-1}}$, the continuity solution
recovers the deterministic preprocessed density over the main-density region. The
volume-weighted recovery errors are small, and the continuity/preprocessor inventory
ratio remains close to unity, as shown in
\cref{tab:ionization_metrics}. This indicates that the face-flux closure can be
extended to a prescribed ionization-loss term for this moderate-loss case.

\begin{table}[!tbp]
    \centering
    \caption{Metrics for the prescribed ionization-loss case.}
    \label{tab:ionization_metrics}
    \small
    \begin{tabular}{lc}
    \toprule
    Quantity & Value \\
    \midrule
    $\nu_{\max}$ & $3.0\times10^{4}\ \mathrm{s^{-1}}$ \\
    $\Phi_{\mathrm{in}}$ & $8.272\times10^{16}\ \mathrm{s^{-1}}$ \\
    $\Phi_{\mathrm{out}}$ & $5.735\times10^{16}\ \mathrm{s^{-1}}$ \\
    $\Phi_{\mathrm{ion}}/\Phi_{\mathrm{in}}$ & $0.3068$ \\
    Recovery $L_1(n)$ & $8.01\times10^{-7}$ \\
    Recovery $L_2(n)$ & $6.19\times10^{-6}$ \\
    Continuity/preprocessor inventory ratio & $1.000001$ \\
    \bottomrule
\end{tabular}

\end{table}

The reported ionization calculation uses the same deterministic DFEM spatial
discretization and velocity-space resolution as the baseline deterministic calculation
and does not rely on a conservative face-flux projection.
Stronger localized absorption cases may require a more robust positivity-preserving and
balance-preserving handoff strategy. The present case is therefore interpreted as a
controlled moderate-loss verification of the ionization-modified $S_N$ preprocessor, its
comparison with the particle-based FM reference, and the extension of the reduced
continuity closure to a prescribed volumetric neutral sink.

% =========================
% 4. Conclusions
% =========================
\section{Conclusions}

This work focused on the closure of reduced neutral-continuity equations for
Hall-thruster simulations in the low-pressure free-molecular limit. Rather than
prescribing an empirical neutral velocity field, the neutral transport problem was first
solved as a free-molecular preprocessing step, so that the reference density field and the
mean-velocity or face-normal-flux quantities needed to close the continuity equation were
obtained from the same transport model. A particle-based free-molecular face-flux
preprocessor was used as a stochastic reference for this handoff, and an $S_N$-DFEM
deterministic preprocessor was developed to generate the corresponding reference density,
velocity moments, and face-normal fluxes under the same inlet, open-boundary, and mixed
diffuse/specular wall-reflection models.

The main findings are summarized as follows:
\begin{itemize}
\item The $S_N$-DFEM preprocessor preserves the dominant neutral-density and velocity
      structures obtained from the particle-based free-molecular reference, while avoiding
      particle sampling noise in the reconstructed moments and face-normal fluxes.
\item In the baseline continuity-recovery test, the deterministic face-flux handoff
      recovers the preprocessed density much more accurately than the particle-based
      closure. The statistical error associated with the face-flux closure is reduced by
      about three orders of magnitude, confirming the consistency of the deterministic
      preprocessor at the reduced-continuity level.
\item The velocity-space resolution study shows that, for the baseline mixed-reflection
      case, the density field is more sensitive to speed-group resolution than to further
      angular refinement once a sufficient angular resolution is used. The
      $N_\Omega=400$, $N_g=16$ setting provides a practical working resolution for the
      parameter studies considered here.
\item The wall-reflection scan shows that near-specular reflection is the most challenging
      regime for the $S_N$ discretization. Specular reflection preserves directional
      memory and strengthens ray-like angular-quadrature artifacts, whereas diffuse
      reflection redistributes the reflected flux and weakens directional coherence.
\item The cylindrical-geometry diagnostic confirms that part of the radial density
      asymmetry in the $r$-$z$ density maps arises from annular area weighting rather than
      from numerical inconsistency. The inlet-density, axial-drift, and inlet-width scans
      further show that the preprocessor captures the distinct effects of source strength,
      residence time, source footprint, and throughput on the free-molecular neutral field.
\item A prescribed moderate ionization-loss case demonstrates that volumetric neutral
      removal can be incorporated into both the deterministic preprocessor and the reduced
      continuity handoff when the same loss frequency is used.
\end{itemize}

The present model remains a reduced axisymmetric free-molecular treatment. The velocity
directions are restricted to the meridional $r$-$z$ plane, the diffuse reflection
fraction is prescribed through a Maxwell-type wall model, and the ionization-loss case is
not self-consistently coupled to a plasma solver. Nevertheless, the results show that an
$S_N$-DFEM free-molecular preprocessor can provide a low-noise, face-flux-consistent
closure for reduced neutral-continuity models. Future work will focus on coupling this
closure to plasma solvers, improving positivity- and balance-preserving handoff
strategies for stronger localized absorption, extending the velocity-space treatment, and
including ion-neutral charge-exchange and momentum-exchange source terms.

% =========================
% Acknowledgments
% =========================
\section*{Acknowledgments}

The authors acknowledge the support from the National Natural Science Foundation of China
(Grant No. 52472403).

\appendix

\section{Supplementary details for flux-consistent \texorpdfstring{\(S_N\)}{SN}-DFEM preprocessing}
\label{app:sn_details}

This appendix retains only the supplementary definitions that affect the deterministic
free-molecular preprocessor and its face-flux handoff. Standard aspects of the
discrete-ordinates angular quadrature and the upwind DFEM sweep are not repeated. The
details included here are those needed to reproduce the present neutral closure:
speed-group quadrature and inlet-flux normalization, mixed diffuse/specular
wall-reflection normalization, the boundary-coupled fixed-point iteration induced by wall
reflection, and the reconstruction of face-normal fluxes from the converged angular-flux
solution.

\subsection{Speed-group quadrature and inlet-flux normalization}
\label{appsubsec:speed_inlet}

Although this is equivalent to an energy grouping for a fixed neutral species, the
speed-group notation is used here because free-molecular residence time, face fluxes, and
ionization attenuation depend directly on molecular speed. The two-dimensional
velocity-space measure is
$v\,\dd v\,\dd\theta$, so the speed group weight is
\begin{equation}
    W_g
    =
    \int_{v_{g-\frac{1}{2}}}^{v_{g+\frac{1}{2}}}
    v\,\dd v
    =
    \frac{1}{2}
    \left(
    v_{g+\frac{1}{2}}^2
    -
    v_{g-\frac{1}{2}}^2
    \right),
    \label{eq:sn_speed_weight}
\end{equation}
where $v_{g-\frac{1}{2}}$ and $v_{g+\frac{1}{2}}$ are the group edges. With angular
weight $w_a^\Omega$, the combined speed-angle quadrature weight is
\begin{equation}
    w_{g,a}=W_g w_a^\Omega .
    \label{eq:sn_combined_weight}
\end{equation}

The inlet source is imposed on a prescribed radial band on the lower axial boundary. The
incoming inlet ordinates satisfy $\eta_a>0$. With
\begin{equation}
    \sigma_{\mathrm{in}}
    =
    \sqrt{\frac{k_B T_{\mathrm{in}}}{m_n}},
    \qquad
    s=\frac{u_{z0}}{\sigma_{\mathrm{in}}},
    \label{eq:sn_sigma_in}
\end{equation}
the positive axial mean of the drifted Maxwellian is
\begin{equation}
    \left\langle u_z^+\right\rangle
    =
    u_{z0}
    +
    \sigma_{\mathrm{in}}
    \frac{\phi(s)}{\Phi(s)},
    \label{eq:sn_positive_mean}
\end{equation}
where $\phi$ and $\Phi$ are the standard normal probability density and cumulative
distribution functions. The prescribed inlet number-flux density is therefore
\begin{equation}
    \Gamma_{\mathrm{in}}
    =
    n_{\mathrm{in}}
    \left\langle u_z^+\right\rangle .
    \label{eq:sn_inlet_flux}
\end{equation}

The discrete incoming distribution is written as
\begin{equation}
    \psi_{g,a}^{\mathrm{in}}
    =
    C_{\mathrm{in}}G_{g,a}^{\mathrm{in}},
    \qquad
    \eta_a>0,
    \label{eq:sn_inlet_distribution}
\end{equation}
where $G_{g,a}^{\mathrm{in}}$ gives the unnormalized speed-angle shape. The normalization
constant enforces the half-range inlet flux,
\begin{equation}
    C_{\mathrm{in}}
    =
    \frac{
    \Gamma_{\mathrm{in}}
    }{
    \displaystyle
    \sum_{g=1}^{N_g}
    \sum_{\eta_a>0}
    w_{g,a}v_g\eta_a
    G_{g,a}^{\mathrm{in}}
    } .
    \label{eq:sn_inlet_C}
\end{equation}
In the baseline calculations, $G_{g,a}^{\mathrm{in}}$ is selected to match the
particle-based inlet sampling. With $u_r=v_g\mu_a$ and $u_z=v_g\eta_a$, the unnormalized
shape is
\begin{equation}
    G_{g,a}^{\mathrm{in}}
    =
    \frac{
    \exp
    \left[
    -\frac{1}{2}
    \left(
    \frac{u_r^2}{\sigma_{\mathrm{in}}^2}
    +
    \frac{(u_z-u_{z0})^2}{\sigma_{\mathrm{in}}^2}
    \right)
    \right]
    }{
    \max(u_z,u_{\min})
    },
    \qquad
    \eta_a>0 .
    \label{eq:sn_inlet_shape}
\end{equation}
The division by the positive axial speed converts the sampled crossing distribution into
a phase-space density; the normalization in Eq.~\eqref{eq:sn_inlet_C} then sets the
prescribed number flux.

\subsection{Mixed diffuse/specular wall-reflection boundary condition}
\label{appsubsec:wall_bc}

The wall boundary condition is retained here because it is one of the main differences
between the present neutral free-molecular problem and conventional neutron-transport
applications. Instead of only vacuum or purely reflecting boundaries, the Hall-thruster
neutral model requires a mixed diffuse/specular gas-surface reflection law. At a wall
face, outgoing ordinates satisfy $\OmegaVec_a\cdot\vect{n}>0$ and incoming ordinates
satisfy $\OmegaVec_a\cdot\vect{n}<0$, where $\vect{n}$ is the outward normal of the
gas-side cell. Open boundaries use zero incoming trace,
\begin{equation}
    \psi_{g,a}^{\mathrm{in}}=0,
    \qquad
    \OmegaVec_a\cdot\vect{n}<0 .
    \label{eq:sn_open_boundary}
\end{equation}

For a wall face, the reflected incoming trace is
\begin{equation}
    \psi_{g,a}^{\mathrm{in},w}
    =
    (1-\alpha_d)
    \mathcal{R}_{\mathrm{sp}}
    \left[
    \psi^{\mathrm{out},w}
    \right]_{g,a}
    +
    \alpha_d
    \mathcal{R}_{\mathrm{df}}
    \left[
    \psi^{\mathrm{out},w}
    \right]_{g,a},
    \qquad
    \OmegaVec_a\cdot\vect{n}<0 ,
    \label{eq:sn_mixed_wall_operator}
\end{equation}
where $\alpha_d$ is the diffuse reflection fraction. For the specular component, the
outgoing preimage of incoming ordinate $a$ is approximated by the nearest outgoing
angular node in the same speed group,
\begin{equation}
    a_R
    =
    \arg\max_{b:\OmegaVec_b\cdot\vect{n}>0}
    \left[
    \OmegaVec_b\cdot
    \left(
    \OmegaVec_a
    -
    2
    \left(
    \OmegaVec_a\cdot\vect{n}
    \right)
    \vect{n}
    \right)
    \right],
    \qquad
    \mathcal{R}_{\mathrm{sp}}
    \left[
    \psi^{\mathrm{out},w}
    \right]_{g,a}
    =
    \psi_{g,a_R}^{\mathrm{out},w}.
    \label{eq:specular_operator}
\end{equation}

For the diffuse component, only the total outgoing number-flux density incident on the
wall is retained,
\begin{equation}
    \Gamma_w^{\mathrm{out}}
    =
    \sum_{g'=1}^{N_g}
    \sum_{\OmegaVec_b\cdot\vect{n}>0}
    w_{g',b}v_{g'}
    \left(
    \OmegaVec_b\cdot\vect{n}
    \right)
    \psi_{g',b}^{\mathrm{out},w}.
    \label{eq:wall_outgoing_flux}
\end{equation}
The wall-temperature speed shape is
\begin{equation}
    M_w(v_g)
    =
    \exp\left[
    -\frac{1}{2}
    \left(
    \frac{v_g}{\sigma_w}
    \right)^2
    \right],
    \qquad
    \sigma_w=
    \sqrt{\frac{k_B T_w}{m_n}} .
    \label{eq:sn_wall_shape}
\end{equation}
The diffuse operator is normalized so that the outgoing wall flux is re-emitted into the
incoming half-space:
\begin{equation}
    \mathcal{R}_{\mathrm{df}}
    \left[
    \psi^{\mathrm{out},w}
    \right]_{g,a}
    =
    \frac{
    M_w(v_g)\Gamma_w^{\mathrm{out}}
    }{
    \displaystyle
    \sum_{g''=1}^{N_g}
    \sum_{\OmegaVec_c\cdot\vect{n}<0}
    w_{g'',c}v_{g''}
    \left|
    \OmegaVec_c\cdot\vect{n}
    \right|
    M_w(v_{g''})
    },
    \qquad
    \OmegaVec_a\cdot\vect{n}<0 .
    \label{eq:diffuse_operator}
\end{equation}
Thus, the specular part preserves directional memory, while the diffuse part integrates
the incident wall flux and redistributes it over all incoming ordinates according to the
wall-temperature speed shape.

\subsection{Boundary-coupled fixed-point iteration}
\label{appsubsec:wall_iteration}

The iteration used here is not a volume-scattering source iteration. It is a fixed-point
procedure for the wall-reflection boundary coupling: the incoming wall trace for the next
sweep depends on the outgoing wall trace obtained from the previous angular solution.
Let $\psi^{(s)}$ denote the angular solution at iteration $s$. For each wall face, the
incoming trace used during the next sweep is
\begin{equation}
    \begin{aligned}
    \psi_{g,a}^{\mathrm{in},w,(s)}
    &=
    (1-\alpha_d)
    \psi_{g,a_R}^{\mathrm{out},w,(s)}
    \\
    &\quad+
    \alpha_d
    \frac{
    M_w(v_g)\Gamma_w^{\mathrm{out},(s)}
    }{
    \displaystyle
    \sum_{g'=1}^{N_g}
    \sum_{\OmegaVec_b\cdot\vect{n}<0}
    w_{g',b}v_{g'}
    \left|
    \OmegaVec_b\cdot\vect{n}
    \right|
    M_w(v_{g'})
    },
    \qquad
    \OmegaVec_a\cdot\vect{n}<0 ,
    \end{aligned}
    \label{eq:wall_incoming_iter}
\end{equation}
where $\Gamma_w^{\mathrm{out},(s)}$ is evaluated from Eq.~\eqref{eq:wall_outgoing_flux}
using $\psi^{(s)}$.

The computational workflow is: prescribe inlet and open-boundary traces, construct the
wall incoming trace from the previous iterate, perform an upwind DFEM sweep for each
speed-angle state, update outgoing wall traces, and repeat until convergence. The
relative change is measured over all local DG degrees of freedom,
\begin{equation}
    \epsilon^{(s+1)}
    =
    \frac{
    \left\|
    \psi^{(s+1)}
    -
    \psi^{(s)}
    \right\|_1
    }{
    \max
    \left(
    \left\|
    \psi^{(s)}
    \right\|_1,
    1
    \right)
    } .
    \label{eq:sn_source_residual}
\end{equation}
The iteration stops when $\epsilon^{(s+1)}\le\epsilon_{\mathrm{tol}}$ or when the
maximum number of iterations is reached.

\subsection{Face-normal flux reconstruction for continuity handoff}
\label{appsubsec:face_flux_reconstruction}

After the boundary-coupled iteration converges, the angular DFEM solution is integrated
over the discrete velocity space to obtain the reference density and mean velocity. If
$\overline{\psi}_{g,a,i,k}$ denotes the volume average in cell $(i,k)$, then
\begin{equation}
    n_{i,k}
    =
    \sum_{g=1}^{N_g}
    \sum_{a=1}^{N_\Omega}
    w_{g,a}
    \overline{\psi}_{g,a,i,k},
    \label{eq:sn_density_app}
\end{equation}
\begin{equation}
    \Gamma_{r,i,k}
    =
    \sum_{g=1}^{N_g}
    \sum_{a=1}^{N_\Omega}
    w_{g,a}v_g\mu_a
    \overline{\psi}_{g,a,i,k},
    \qquad
    \Gamma_{z,i,k}
    =
    \sum_{g=1}^{N_g}
    \sum_{a=1}^{N_\Omega}
    w_{g,a}v_g\eta_a
    \overline{\psi}_{g,a,i,k}.
    \label{eq:sn_flux_components_app}
\end{equation}
The mean neutral velocity is $u_{r,i,k}=\Gamma_{r,i,k}/n_{i,k}$ and
$u_{z,i,k}=\Gamma_{z,i,k}/n_{i,k}$ for cells with $n_{i,k}>0$.

The continuity handoff uses signed face-normal fluxes rather than only cell-centered
velocity moments. The sign convention is that a positive radial face flux points from
the lower-$r$ cell to the higher-$r$ cell, and a positive axial face flux points from the
lower-$z$ cell to the higher-$z$ cell. For an internal radial face between a left cell
$L$ and a right cell $R$, the upwind face flux is
\begin{equation}
    \Gamma_{r,f}
    =
    \sum_{g=1}^{N_g}
    \sum_{\mu_a>0}
    w_{g,a}v_g\mu_a
    \psi_{g,a,L}^{f}
    +
    \sum_{g=1}^{N_g}
    \sum_{\mu_a<0}
    w_{g,a}v_g\mu_a
    \psi_{g,a,R}^{f}.
    \label{eq:sn_internal_r_flux}
\end{equation}
For an internal axial face between a lower cell $B$ and an upper cell $T$,
\begin{equation}
    \Gamma_{z,f}
    =
    \sum_{g=1}^{N_g}
    \sum_{\eta_a>0}
    w_{g,a}v_g\eta_a
    \psi_{g,a,B}^{f}
    +
    \sum_{g=1}^{N_g}
    \sum_{\eta_a<0}
    w_{g,a}v_g\eta_a
    \psi_{g,a,T}^{f}.
    \label{eq:sn_internal_z_flux}
\end{equation}
At an open boundary, only outgoing states from the interior trace are included; incoming
states from outside the domain are zero by Eq.~\eqref{eq:sn_open_boundary}. These
face-normal fluxes are the deterministic counterpart of the particle face-crossing
statistics and are used as $\Gamma_f^{\mathrm{ref}}$ in the continuity-recovery relation
$F_f(n^{\mathrm{ref}})=\Gamma_f^{\mathrm{ref}}$.

% =========================
% References
% =========================
\bibliographystyle{elsarticle-num}
\bibliography{paper1_reference}

\end{document}